\documentclass[sigconf,authorversion,natbib=true,nonacm]{acmart}

\AtBeginDocument{%
  }

\setcopyright{acmcopyright}
\copyrightyear{2023}
\acmYear{2023}
\acmDOI{10.1145/3604237.3626884}

\acmConference[ICAIF-2023]{ICAIF-2023: ACM International Conference on AI in Finance}{ 27-29 November, 2023}{New York, NY}

\usepackage{multirow}
\usepackage{amsmath}
\usepackage{relsize}
\usepackage{graphicx}
\usepackage{listings}
\usepackage{xcolor}
\usepackage{booktabs}
\usepackage{fancyref}
\usepackage{todonotes}
\usepackage{import}
\usepackage{dsfont}
\usepackage{lineno}
\usepackage{float}
\usepackage{caption}
\usepackage[skip=10pt]{subcaption}
\usepackage{bm}
\usepackage{bbm}
\usepackage{commath}
\usepackage[section]{placeins}
\usepackage{fancyhdr}
\usepackage{natbib}
\usepackage{datetime}

\DeclareMathOperator*{\argmax}{argmax}

\def\RR{\mathbb{R}}

\usepackage{algorithmic}
\usepackage[ruled,vlined]{algorithm2e}
\usepackage{makecell}
\usepackage{ifthen}
\usepackage{makecell}
\usepackage{siunitx}

\usepackage{amsthm}

\usepackage{varwidth}
\usepackage{capt-of}

\usepackage{tikz}
\usetikzlibrary{calc,matrix,chains,positioning,decorations.pathreplacing,arrows}

\newdate{date}{1}{1}{2023}
\newcommand{\myref}[2]{\hyperref[#2]{#1 \ref*{#2}}}

\newcommand{\na}{\textcolor{gray}{\footnotesize N/A}}

\theoremstyle{definition} 
\begin{document}

\title{Generative Machine Learning for Multivariate Equity Returns}

\author{Ruslan Tepelyan}
\authornote{Both authors contributed equally to this research.}
\affiliation{%
  \institution{Bloomberg}
  \streetaddress{120 Park Ave}
  \city{New York}
  \country{USA}}
\email{rtepelyan@bloomberg.net}

\author{Achintya Gopal\footnotemark[1]{}}
\affiliation{%
  \institution{Bloomberg}
  \streetaddress{731 Lexington Ave}
  \city{New York}
  \country{USA}}
\email{agopal6@bloomberg.net}



\begin{abstract}

The use of machine learning to generate synthetic data has grown in popularity with the proliferation of text-to-image models and especially large language models. The core methodology these models use is to learn the distribution of the underlying data, similar to the classical methods common in finance of fitting statistical models to data. In this work, we explore the efficacy of using modern machine learning methods, specifically conditional importance weighted autoencoders (a variant of variational autoencoders) and conditional normalizing flows, for the task of modeling the returns of equities. The main problem we work to address is modeling the joint distribution of all the members of the S\&P 500, or, in other words, \textit{learning a 500-dimensional joint distribution}. We show that this generative model has a broad range of applications in finance, including generating realistic synthetic data, volatility and correlation estimation, risk analysis (e.g., value at risk, or VaR, of portfolios), and portfolio optimization. 
\end{abstract}

\maketitle

\section{Introduction}

Risk forecasting is an important problem in finance to help estimate portfolio risk, volatility, and correlation, as well as being an input for portfolio optimization. These problems are often solved via statistical quantities such as quantiles (for VaR), standard deviation (for volatility), Pearson correlations, etc.
One approach is to derive an estimator specific to the statistical quantity (e.g., \cite{vol_ohlc_1980,ledoit2022correstimator}); another approach is to develop a statistical model such as GARCH \citep{Engle1982ARCH} or multivariate GARCH \citep{engle2002dcc} that learns the distribution of returns, after which any statistical quantity can be queried from the learned distribution. 
In this work, we focus on a deep learning approach to developing a statistical model that learns the multivariate distribution of stock returns.

The generative models of deep learning are also just statistical models; examples include variational autoencoders (VAEs, \cite{kingma2013vae}), normalizing flows \citep{dinh2014nice,Rezende2015NF}, autoregressive models as used in large language models \citep{vaswani2017attention,radford2019language}, etc. Given this, we ask the question, ``can we use deep probabilistic models for the task of modeling distributions of returns?''
Phrased another way, the problem is: given all of the data up to a particular point in time, can we  model the return of a stock over the next day as an arbitrary probability distribution? 
Modern deep learning techniques allow us to fit such distributions, and as long as these distributions are well-calibrated, we can rely, in a statistical sense, on any downstream results derived from them.

Prior work has applied deep learning approaches to model single financial time series (e.g., \cite{Buhler2020Generate, wiese2019quantgan}); \citet{Wiese2021Multi} applied their approach to modeling two assets by modeling each independently first and then correlating them. However, to our knowledge, no prior work has constructed a method to model the joint distribution of an arbitrary number of stocks; without modeling the joint distribution, the generative model cannot capture the correlations between stock returns.
While previous deep learning work has modeled high-dimensional data (e.g., \citep{Oord2016PixelRNN,radford2019language}), 
these methods do not handle a variable number of dimensions, something that is required in modeling stocks, since stocks can be added and removed from the universe (i.e., a varying number of dimensions at each time step).
In this paper, we fill these two gaps. 

Our approach entails building two models that condition on the past: a conditional importance-weighted autoencoder (CIWAE, \citep{burda2015iwae,ESGImp2020}) for the factors (e.g., broad market indices, sector-level indices) and a conditional normalizing flow \citep{winkler2020cflow} for individual stock returns. For sampling, we use an autoregressive approach \citep{Oord2016PixelRNN}. More detail of our model is given in \myref{Section}{sec:methodology}. 
To build these models, we used many factors and features; we ablate our modeling decisions in \myref{Section}{sec:ablation}, offering guidance to future research.

We empirically illustrate our method on the (point-in-time) constituents of the S\&P 500 (\myref{Section}{sec:eval}), or in other words, \textit{we model a near 500-dimensional conditional joint distribution}.
We demonstrate the efficacy of our model across multiple downstream tasks. Specifically, we show that our model outperforms GARCH in the univariate case leading to accurate risk estimations for individual securities (\myref{Section}{sec:risk_eval}), outperforms a classical factor model in the multivariate case leading to accurate risk estimation for portfolios (\myref{Section}{sec:risk_eval}), and, finally, creates a portfolio that outperforms the market (\myref{Section}{sec:portfolio}).
Our results provide evidence that neural networks are a competitive alternative to classical estimators and methods, and we hope these results spur future exploration into the usage of deep generative methods for finance.

\section{Background}

The crux of our methodology is modeling conditional distributions $p(\mathbf{y}|\mathbf{x})$. To do so, we use a conditional importance-weighted autoencoder (CIWAE  \citep{burda2015iwae,ESGImp2020}) and a conditional normalizing flow \citep{winkler2020cflow}. Both generative models allow us to compute log-likelihoods, a task which is both useful for statistical evaluation and for training using maximum likelihood estimation; further, both are generative models, allowing us to evaluate the quality of the synthetic data by sampling from them.

The power of VAEs and normalizing flows over a family of distributions such as a multivariate Gaussian distribution or Normal Inverse Gaussian (\myref{Section}{sec:nig}) is that it can handle not only location, scale and covariance, but also also learn heavy tails, skew, multi-modality, and higher order moments across dimensions.

\subsection{Normal Inverse Gaussian}\label{sec:nig}

The density function of a Normal Inverse Gaussian is:
$$f(x) =  \frac{\alpha \delta K_1\left( \alpha \sqrt{\delta ^2  + (x-\mu)^2} \right)}{\pi \sqrt{\delta ^2 + (x-\mu)^2}} e^{\delta \gamma  + \beta (x - \mu)} $$
where $K_1$ is a modified Bessel function of the second kind, $\gamma=\sqrt{\alpha^2 + \beta^2}$, $\gamma, \delta > 0$ and $\mu, \beta \in \RR $.
The power of Normal Inverse Gaussians is that they are able to the model the location $\mu$, scale $\delta$, heavy tailedness $\alpha$, and asymmetry (i.e., skew) $\beta$.

\subsection{VAE, CVAE, and CIWAE}\label{sec:vae}

Suppose that we wish to formulate a joint distribution on an $n$-dimensional real vector $x$. For a VAE-based approach, the generative process is defined as:
$$ \mathbf{x} \sim p(\mathbf{x} | \mathbf{z}) \qquad \mathbf{z} \sim p(\mathbf{z}) $$
where $\mathbf{x}$ is the observed data and $\mathbf{z}$ is a latent variable.
VAEs use a neural network to parameterize the distribution $p(\mathbf{x} | \mathbf{z})$. 

Say we are modeling $p(y|x)$, we can change the generative process to:
$$ \mathbf{y} \sim p(\mathbf{y} | \mathbf{z}, \mathbf{x}) \qquad \mathbf{z} \sim p(\mathbf{z}) $$
For the generative process, we can see sampling from a VAE requires \textit{two sampling steps}: sampling from $p(\mathbf{z})$ and then sampling from $p(\mathbf{y} | \mathbf{z}, \mathbf{x})$. Often $p(\mathbf{y} | \mathbf{z}, \mathbf{x})$ is referred to as the decoder.

When fitting distributions to data, the most common method is maximum likelihood estimation (MLE):
$$ \argmax_{\theta} \sum_{i=1}^{N} \log p(\mathbf{y_i} | \mathbf{x_i}, \theta) $$
where $\{\mathbf{x_i},\mathbf{y_i}\}_{i=1}^{N}$ denotes the data we want to fit our model on. For a latent variable model, the log-likelihood is:
 $$ \log p(\mathbf{y_i} | \mathbf{x_i}, \theta) = \log \int p(\mathbf{x_i} | \mathbf{x_i}, \mathbf{z}, \theta) p(\mathbf{z}) dz$$
This integral is the reason we cannot directly perform gradient descent of the total log-likelihood; VAEs instead use variational inference \citep{kingma2013vae}, i.e, optimize the evidence lower bound (ELBo). 
Since we are modeling the conditional distribution, we give the conditional version of the VAE loss:
\begin{align}
 \log p(\mathbf{y_i} | \mathbf{x_i}, \theta) &\geq \mathbb{E}_{\mathbf{z} \sim q(\mathbf{z} | \mathbf{x_i}, \mathbf{y_i})}\left[\log  \frac{p(\mathbf{y_i} |\mathbf{x_i}, \mathbf{z}, \theta)p(\mathbf{z} | \mathbf{x_i}, \theta) }{q(\mathbf{z} | \mathbf{y_i}, \mathbf{x_i})} \right] 
 \\ &= -\mathcal{L}_{\text{CVAE}}(\mathbf{y_i}, \mathbf{x_i}; \theta, \phi) 
\end{align}

To be able to train VAEs with gradient descent, we need to be able to differentiate through the sampling of $z$ from $q(\mathbf{z} | \mathbf{x_i}, \mathbf{y_i})$ (often referred to as the encoder).  To achieve this, \citet{kingma2013vae} use the reparameterization trick: a way to reparametrize sampling from a distribution by ensuring the operation with randomness is not dependent on quantities for which we want to differentiate with respect to. For example, to sample $z$ from a normal distribution $\mathcal{N}(\mu, \sigma)$:
$$ z = \mu +  \sigma \epsilon \qquad \epsilon \sim \mathcal{N}(0, 1)$$

A simple trick to help close the gap in the lower bound is to use importance weighted autoencoders (IWAE). 
Similar to CVAE, we give the conditional version of the IWAE loss:
\begin{align}
 \log p(\mathbf{y_i} | \mathbf{x_i}, \theta) &\geq \mathbb{E}_{\mathbf{z_1}, \dots, \mathbf{z_k} \sim q(\mathbf{z} | \mathbf{x_i}, \mathbf{y_i})}\left[\log  \sum_{j=1}^{k}  \frac{p(\mathbf{y_i} |\mathbf{x_i}, \mathbf{z_j}, \theta)p(\mathbf{z_j} | \mathbf{x_i}, \theta) }{q(\mathbf{z_j} | \mathbf{y_i}, \mathbf{x_i})} \right] 
 \\ &= -\mathcal{L}_{\text{CIWAE}, k}(\mathbf{y_i}, \mathbf{x_i}; \theta, \phi) 
\end{align}
\citet{burda2015iwae} showed that increasing $k$ reduces the bias of the estimator.

\subsection{Normalizing Flows}\label{sec:normalizing_flows}

Suppose that we wish to formulate a joint distribution on an $n$-dimensional real vector $x$. 
The generative process for flows is defined as:
$$ \mathbf{x} = g(\mathbf{z}) \qquad \mathbf{z} \sim p_z(\mathbf{z}) $$
where $p_z$ is often a normal distribution and $g$ is an invertible function. Let $f = g^{-1}$. Using change of variables, the log-likelihood of $\mathbf{x}$ is
$$ \log p_x(\mathbf{x}) = \log p_z\left (f(\mathbf{x}) \right) + \log\ \abs{\text{det}\left(\frac{\partial f(\mathbf{x})}{\partial \mathbf{x}}\right)} $$
To train flows (i.e., maximize the log-likelihood of data points), we need to be able to compute the logarithm of the absolute value of the determinant of the Jacobian of $f$, also called the \textit{log-determinant}.

To construct expressive normalizing flows, we can compose many less expressive ones ($g_i$) as this is still invertible, differentiable, and tractable since the log-determinant of this composition is the sum of the individual log-determinants. Thus, in theory, a potentially complex transformation can be built up from a series of smaller, simpler transformations with tractable log-determinants \citep{CFINN2020}. 

Constructing a normalizing flow model in this way provides two obvious applications: drawing samples using the generative process and evaluating the probability density of the modeled distribution by computing $p_x(x)$. These require evaluating the inverse transformation $f$, the log-determinant, and the density $p_z(z)$. In practice, if inverting either $g$ or $f$ turns out to be inefficient, then one or the other of these two applications can become intractable. 

For evaluating the probability density, in particular, computing the log-determinant can be an additional trouble spot.
However, since we only model one-dimensional targets in this work,
the issue of computing the log-determinant is moot; as long as our function $f$ is invertible, then we can compute using automatic differentiation $\abs{\frac{\partial f(x)}{\partial x}}$.

\subsubsection{Residual Flows} 

There are many approaches in the literature to designing a normalizing flow; a thorough literature survey of normalizing flows has been conducted by \citet{papamakarios2021normalizing}. In our work, we use residual flows as prior work \citep{QuARFlows2020} has found residual flows have inductive biases that lead to smoother transformations.

A residual flow is a residual network $\left(f(x) = x + \mathcal{F}(x)\right)$ where the Lipschitz constant of $\mathcal{F}$ is strictly less than one. 
This constraint on the Lipschitz constant ensures invertibility; the transform is invertible using Banach's fixed point algorithm (\myref{Algorithm}{alg:fixed_point_iteration}) where the convergence rate is exponential in the number of iterations and is faster for smaller Lipschitz constants \citep{Behrmann2019}.

\begin{algorithm}[t]
   \caption{Inverse of Residual Flow via Fixed Point Iteration}
   \label{alg:fixed_point_iteration}
\begin{algorithmic}
   \STATE {\bfseries Input:} data $y$, residual block $g$, number of iterations $n$
   \STATE Initialize $x_0 = y$.
   \FOR{$i=1$ {\bfseries to} $n$}
   \STATE $x_i = y - g(x_{i-1})$
   \ENDFOR
\end{algorithmic}
\end{algorithm}

\subsubsection{Conditional Residual Flows}\label{sec:cond_res_flow}

Say we are modeling $p(y|\mathbf{x})$, we can change the generative process to:
$$ \mathbf{y} = g(\mathbf{z}, \mathbf{x}) \qquad \mathbf{z} \sim p_z(\mathbf{z}| \mathbf{x}) $$
%
In order to make the residual flows conditional, we simply concatenate a representation of the features with $y$ in the residual block. More concretely, we first pass the features $x$ into a neural network to get a representation $\mathbf{h}$; then the residual flow will be $\left(f(y) = y + \mathcal{F}([y || \mathbf{h}])\right)$ where  $||$ denotes concatenation. 

\section{Methodology}\label{sec:methodology}

\begin{figure*}[!bt]
  \begin{center}
  \begin{tikzpicture}[
    every neuron/.style={
      circle,
      minimum size=0.3cm,
      very thick
    },
    every data/.style={
      rectangle,
      minimum size=0.4cm,
      thick
    },
  ]
  
    \node [align=center,every neuron/.try, data 1/.try, minimum width=0.3cm] (ema_fac)  at ($ (2.5, 0) $) {$\mathbf{\hat{F}_{E, T + 1}}$};
    \node [align=center,every data/.try, data 1/.try, minimum width=0.3cm,draw] (ema)  at ($ (ema_fac) - (1.5, 0) $) {EMA};
    \node [align=center,every neuron/.try, data 1/.try, minimum width=0.3cm] (input-0)  at ($ (ema) - (1.5, 0) $) {$\mathbf{\hat{F}_{T + 1}}$};
    \node [align=center,every data/.try, data 1/.try, minimum width=0.3cm,draw] (pca)  at ($ (input-0) - (1.5, 0) $) {PCA};
    \node [align=center,every neuron/.try, data 1/.try, minimum width=0.3cm] (input-00)  at ($ (pca) - (1.5, 0) $) {$\mathbf{F_{T+1}}$};
  
    \node [align=center,every neuron/.try, data 1/.try, minimum width=0.3cm] (ft)  at ($ (ema_fac) + (0,1.5) $) {$\mathbf{\hat{F}_{E, T}}$};
    \node [align=center,every neuron/.try, data 1/.try, minimum width=0.3cm] (input-3)  at ($ (ft) + (0,1.5) $) {$\mathbf{\vdots}$};
    \node [align=center,every neuron/.try, data 1/.try, minimum width=0.3cm] (f1)  at ($ (input-3) + (0,1.5) $) {$\mathbf{\hat{F}_{E,1}}$};
  
    \node [align=center,every data/.try, data 1/.try, minimum width=0.3cm, draw] (EMA-t)  at ($ (ft) - (1.5, 0) $) {EMA};
    \node [align=center,every data/.try, data 1/.try, minimum width=0.3cm, draw] (EMA-1)  at ($ (f1) - (1.5, 0) $) {EMA};

    \node [align=center,every neuron/.try, data 1/.try, minimum width=0.3cm] (EMAf-t)  at ($ (EMA-t) - (1.5, 0) $) {$\mathbf{\hat{F}_{T}}$};
    \node [align=center,every neuron/.try, data 1/.try, minimum width=0.3cm] (EMAf-1)  at ($ (EMA-1) - (1.5, 0) $) {$\mathbf{\hat{F}_{1}}$};
    \node [align=center,every neuron/.try, data 1/.try, minimum width=0.3cm] (vdots)  at ($ (EMA-t) + (0,1.5) $) {$\mathbf{\vdots}$};

    \draw [black,solid,->] ($(EMAf-1.east)$) -- ($(EMA-1.west)$);
    \draw [black,solid,->] ($(EMAf-t.east)$) -- ($(EMA-t.west)$);
    \draw [black,solid,->] ($(EMA-t.east)$) -- ($(ft.west)$);
    \draw [black,solid,->] ($(EMA-1.east)$) -- ($(f1.west)$);

    \node [align=center,every data/.try, data 1/.try, minimum width=0.3cm, draw] (PCA-t)  at ($ (EMAf-t) - (1.5, 0) $) {PCA};
    \node [align=center,every data/.try, data 1/.try, minimum width=0.3cm, draw] (PCA-1)  at ($ (EMAf-1) - (1.5, 0) $) {PCA};

    \node [align=center,every neuron/.try, data 1/.try, minimum width=0.3cm] (input-t)  at ($ (PCA-t) - (1.5, 0) $) {$\mathbf{F_{T}}$};
    \node [align=center,every neuron/.try, data 1/.try, minimum width=0.3cm] (input-1)  at ($ (PCA-1) - (1.5, 0) $) {$\mathbf{F_{1}}$};

    \draw [black,solid,->] ($(pca.east)$) -- ($(input-0.west)$);
    \draw [black,solid,->] ($(input-00.east)$) -- ($(pca.west)$);
    \draw [black,solid,->] ($(PCA-1.east)$) -- ($(EMAf-1.west)$);
    \draw [black,solid,->] ($(PCA-t.east)$) -- ($(EMAf-t.west)$);
    \draw [black,solid,->] ($(input-t.east)$) -- ($(PCA-t.west)$);
    \draw [black,solid,->] ($(input-1.east)$) -- ($(PCA-1.west)$);

    \node [align=center,every data/.try, data 1/.try,rotate=90, minimum width=4.0cm, minimum height=0.4cm,draw] (lstm1)  at ($ (input-3) + (1.5,0) $) {$\text{LSTM}_1$};
  
    \node [align=center,every data/.try, data 1/.try, minimum width=0.3cm,draw] (vae)  at ($ (ema_fac) + (1.5, 0) $) {CIWAE};
  
    \node [align=center,every data/.try, data 1/.try, minimum width=0.3cm] (vae_loss)  at ($ (vae) - (0, 1.5) $) {$p (\hat{\mathbf{F}}_{T + 1} |  \{\mathbf{\hat{F}}_{E, t}\}_{t=1}^{T})$};
  
    \node [align=center,every neuron/.try, data 1/.try, minimum width=0.3cm] (fut_ret)  at ($ (vae) + (1.5, 0) $) {$r_{i,{T + 1}}$};
  
    \node [align=center,every neuron/.try, data 1/.try, minimum width=0.3cm] (xt)  at ($ (fut_ret) + (0,0.75) $) {$\mathbf{X_{i, T}}$};
  
    \node [align=center,every neuron/.try, data 1/.try, minimum width=0.3cm] (rt)  at ($ (fut_ret) + (0,1.5) $) {$r_{i,{T}}$};
    \node [align=center,every neuron/.try, data 1/.try, minimum width=0.3cm] (input-4)  at ($ (rt) + (0,1.5) $) {$\mathbf{\vdots}$};
    \node [align=center,every neuron/.try, data 1/.try, minimum width=0.3cm] (r1)  at ($ (input-4) + (0,1.5) $) {$r_{i,1}$};
  
    \node [align=center,every data/.try, data 1/.try,rotate=90, minimum width=4.0cm, minimum height=0.4cm,draw] (lstm2)  at ($ (input-4) + (1.5,0) $) {$\text{LSTM}_2$};
    \node [align=center,every data/.try, data 1/.try, minimum width=0.3cm,draw] (flow)  at ($ (fut_ret) + (1.5, 0) $) {Cond. \\ Flow};
  
    \node [align=center,every data/.try, data 1/.try, minimum width=0.3cm] (flow_loss)  at ($ (flow) - (-1.5, 1.5) $) {$p (r_\mathbf{{i, T + 1}} | \mathbf{\hat{F}_{T + 1}},  \{r_{i, t}\}_{t=1}^{T}, \{\mathbf{\hat{F}_{t}}\}_{t=1}^{T}, \mathbf{X_{i,t}})$};
  
    \draw [black,solid,->] ($(input-0.east)$) -- ($(ema.west)$);
    \draw [black,solid,->] ($(ema.east)$) -- ($(ema_fac.west)$);
    \draw [black,solid,->] ($(input-0.east)$) -- ($(vae_loss.north) - (0.1,0)$);
    \draw [black,solid,->] ($(lstm1.west)$) -- ($(vae.north)$);
    \draw [black,solid,->] ($(vae.south)$) -- ($(vae_loss.north)$);
  
    \draw [black,solid,->] ($(ft.east)$) -- ($(ft.east) + (0.75,0)$);
    \draw [black,solid,->] ($(f1.east)$) -- ($(f1.east) + (0.75,0)$);
  
    \draw [black,solid,->] ($(xt.east)$) -- ($(flow.north west)$);
    \draw [black,solid,->] ($(lstm2.west)$) -- ($(flow.north)$);
    \draw [black,solid,->] ($(flow.south)$) -- ($(flow_loss.north)-(1.5,0)$);
    \draw [black,solid,->] ($(fut_ret.east)$) -- ($(flow_loss.north)-(1.5,0)$);
  
    \draw [black,solid,->] ($(rt.east)$) -- ($(rt.east) + (0.75,0)$);
    \draw [black,solid,->] ($(r1.east)$) -- ($(r1.east) + (0.75,0)$);
  
    \path [black,solid,->] ($(input-0.east)$) edge [bend left=20]  ($(flow.west)$);
    \path [black,solid,->] ($(EMAf-t.east)$) edge [bend left=20]  ($(rt.east) + (0.8,0)$);
    \path [black,solid,->] ($(EMAf-1.east)$) edge [bend left=20]  ($(r1.east) + (0.8,0)$);
  
    \draw[red,thick,dotted] ($(lstm1.south east)+(0.5,0.85)$)  rectangle ($(input-00.south west)-(0.25,0.25)$);
    \node [align=center,data 1/.try, minimum width=1.5cm] (factor-model)  at ($(input-0) + (0.5, 5.4)$) {\textit{Factor Model}};
  
    \node [align=center,data 1/.try, minimum width=1.5cm] (stock-model)  at ($(factor-model) + (7.0, 0)$) {\textit{Stock Model}};
    \draw[red,thick,dotted] ($(stock-model.north east)+(0.25,0.25)$)  rectangle ($(fut_ret.south west)-(0.25,0.25)$);

    \node[inner sep=0pt] (russell) at ($ (stock-model.east) + (3.0,-3.0) $)
  {\includegraphics[width=.3\textwidth]{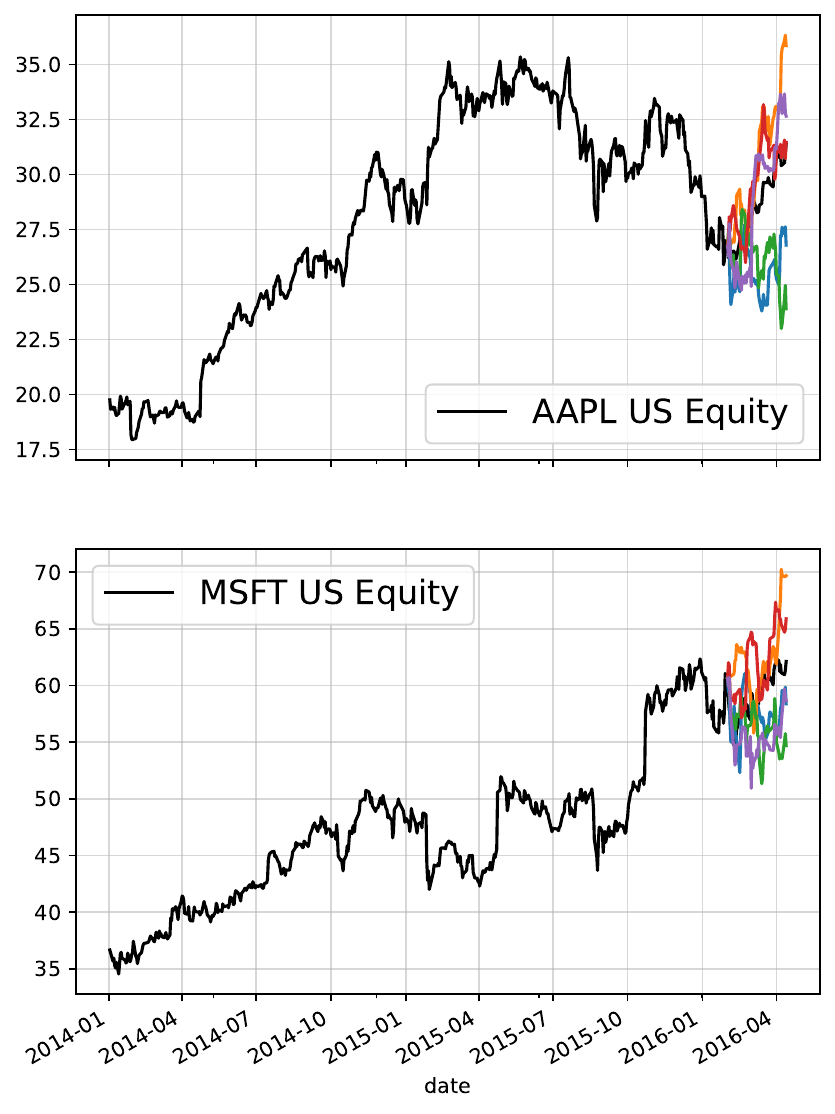}};

  \end{tikzpicture}
  \end{center}
  \caption{We show a high level diagram of our final model architecture. $\mathbf{r_{i, t}}$  denotes the returns of security $i$ at time $t$. $\mathbf{F_{ t}}$ denotes the factor returns at time $t$. $\mathbf{X_{i,t}}$ denotes security level features for  security $i$ at time $t$. We divided the architecture into two pieces: the factor model and the stock model. to the right, we show example samples from our model; the same color represents the same multivariate sample.}\label{fig:final_arch}
  
  \end{figure*}
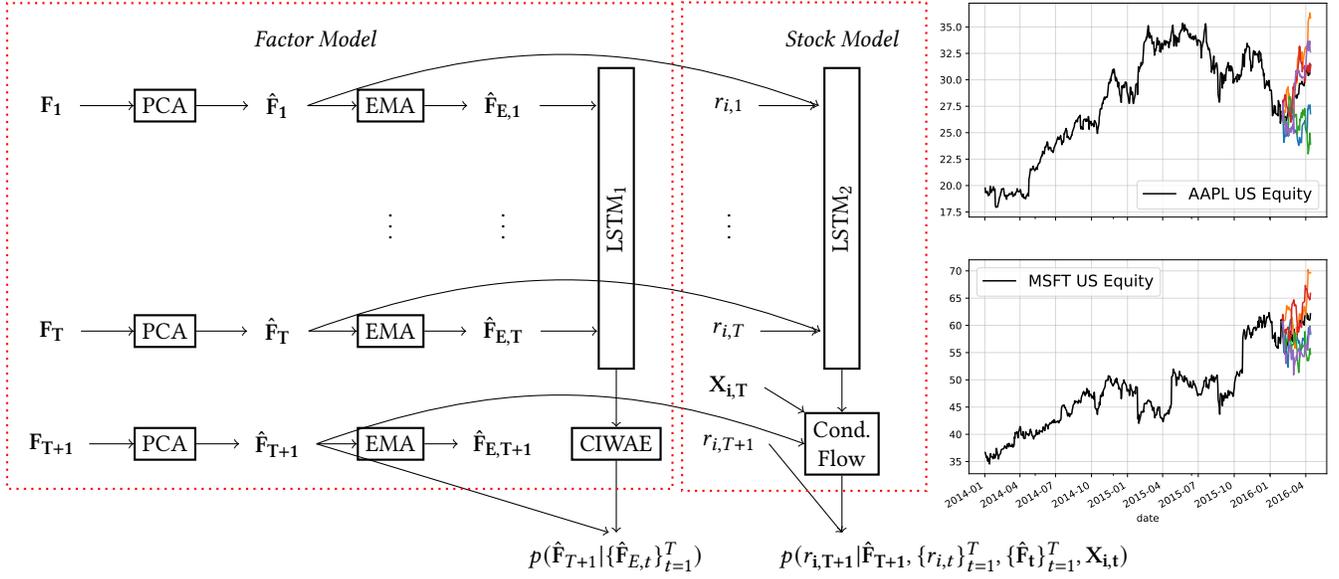

The goal of our model is to learn:
$$ p(\{r_{i,T+1}\}_{i=1}^{N_{T+1}} | \mathcal{F}_T) $$
or, in other words, the joint distribution of $N_{T + 1}$ returns $r_{i, T+1} \in \RR$ at time $T + 1$ given historical data $\mathcal{F}_T$. We use $\mathcal{F}_T$ to represent all the information available up until and including time $T$; the specific information from the set used in our model is defined in \myref{Section}{sec:factor_model} and \myref{Section}{sec:stock_model}. 

To model a variable number of securities, inspired by factor analysis, we rewrite this above likelihood as:
$$\int \left(\prod_{i=1}^{N_{T+1}}  p(r_{i,T+1} | \mathbf{F_{T+1}}, \mathcal{F}_T) \right) p( \mathbf{F_{T+1}} | \mathcal{F}_T) d\mathbf{F_{T+1}}  $$
where $\mathbf{F_{T+1}} \in \RR^{\abs{F}}$ refers to information known at time $T + 1$ that makes the returns of each security $i$ conditionally independent.
We will refer to $\mathbf{F_{T+1}}$ as factors; the specific choice of $\mathbf{F}_{T+1}$ we use for our model will be described in \myref{Section}{sec:factor_model}. 

In \myref{Figure}{fig:final_arch}, we show a high level diagram of our model which has two components: the factor model (\myref{Section}{sec:factor_model}) and the stock model (\myref{Section}{sec:stock_model}).

\subsection{Factor Model}\label{sec:factor_model}

\begin{table}[!bt]
  \centering
  \smaller
  \begin{tabular}{cccc}
\toprule
\textit{Baseline} & \textit{Bond} & \textit{Commodities} & \textit{International}\\
\midrule
VIX  & LBUSTRUU & BCOMINTR & MXJP \\
SKEW &  LUACTRUU &  BCOMAGTR & MXPCJ\\
MOVE & SPUHYBDT  &  BCOMGCTR & MXGB \\ 
RAY & &  MXCN & MXCA \\
FARBAST & & BCOMNGTR & MXMX \\ 
REIT & & BCOMSITR & MXEF  \\
M2US000\$ & & BCOMINTR & MXEUG \\
M1USQU & & & MXBRIC  \\
RAV \\
RAG \\
 DJDVY \\ 
MLCP \\
 MMCP  \\
MSCP \\
\end{tabular}
  \caption{List of indices we used as factors in our model.}
  \label{tab:factors}
\end{table}

Factor analysis is widely used in quantitative finance, and offers many advantages \citep{FamaFrench2015return}. By decomposing stock returns into a linear combination of factor exposures, we can achieve dimensionality reduction and can analyze an ever-changing universe of stocks. Inspired by this approach, we seek to use machine learning techniques to model the arbitrary (nonlinear) dependence of a stock's returns on some set of observed factors. Note that for us, the notion of a factor is along the lines of a macro-economic variable related to financial markets. The full distribution of factors can be thought of as the distribution of market states, or as we will call them in this paper, ``states of the world''. 
This includes broad market indices in the U.S. and other countries, sector-level indices, bond indices, and commodities.
\myref{Table}{tab:factors} contains the list of factors we use; additionally, not listed are all the S\&P GICS Level 2, 3, and 4 indices we used. 
In \myref{Section}{sec:ablation_factors}, we discuss why it is important to use all these indices.

However, it is important that we are thoughtful about our decision on the list of factors. Intuitively, in the most extreme case, if we put all the individual security returns in our list of factors, then $p(r_{i,T+1} | \mathbf{F_{T+1}}, \mathcal{F}_T)$ would be trivial to learn but $p (\mathbf{F_{T+1}}|  \mathcal{F}_T)$ would be extremely difficult (since all the difficulty of forecasting returns has been pushed into our factor model). This implies we need to define a universe of factors that both make the security returns conditionally independent and $p (\mathbf{F_{T+1}}|  \mathcal{F}_T)$ simple to learn.

To simplify learning $p (\mathbf{F_{T+1}}|  \mathcal{F}_T)$, we reduce the dimensionality of our factors by using PCA.
Specifically, we fit a PCA model on the training period and tune the number of components using the joint log-likelihood on the validation set (\myref{Section}{sec:eval}).
Thus, instead of learning $p (\mathbf{F_{T+1}} | \mathcal{F}_T)$, we learn $ p (\mathbf{\hat{F}_{T + 1}} |  \mathcal{F}_T)$ where $\hat{F}_{T + 1}$ refers to the factor returns after applying a PCA transformation. 

For $\mathcal{F}_T$, our factor model looks at the set of PCA transformed factor values in the lookback period $\{\mathbf{\hat{F}_{t}}\}_{t=1}^{T}$  and constructs a distribution over $\mathbf{\hat{F}_{T + 1}}$. Mathematically, our factor model is learning:
$$ p (\mathbf{\hat{F}_{T + 1}} |  \{\mathbf{\hat{F}_{t}}\}_{t=1}^{T})$$

\subsubsection{EMA Features} For a baseline, we use an exponential moving average (EMA) of $\{\mathbf{\hat{F}_{t}}\}_{t=1}^{T}$ to compute a rolling mean $\mathbf{\mu_{F, T}}$ and covariance matrix $\mathbf{\Sigma_{F, T}}$ and use these values to create a multivariate Gaussian.
Further, we use these values in our VAE by transforming $\mathbf{\hat{F}_{T + 1}}$ into:
$$\mathbf{\hat{F}_{E, T + 1}} = \mathbf{\Sigma_{F, T}}^{-1}  \left(\mathbf{\hat{F}_{T + 1}} - \mathbf{\mu_{F, T}}\right) $$
While a neural network should be able to learn this transformation, we find in practice modeling $\mathbf{\hat{F}_{E, T + 1}}$ improves generalization (\myref{Section}{sec:ablate_factor_model}). The intuition is using $\mathbf{\hat{F}_{E, T + 1}}$ implies the VAE simply needs to learn higher order moments.

To ensure a well-designed baseline, as well as strong features, we use a different decay rate in the EMA for each security, independently for the mean and variance. Further, for each EMA, we incorporate two parameters: a shrinkage factor and a value to shrink towards. Thus, for the mean and variance, we have a total of $3N$ parameters to learn (if we have $N$ PCA-components). For the correlation matrix, we also have a single decay rate, a single shrinkage factor, and a correlation matrix to shrink towards ($ \frac{N (N - 1)}{2}$ additional parameters). To optimize all these parameters, we use the mean and covariance matrix to parametrize a multivariate normal and use gradient descent to optimize the negative log-likelihood (NLL) of the observed $\hat{F}_{T}$ values in the training set. To regularize, we add a hyperparameter the penalizes the deviation of each multivariate normal from the multivariate normal parameterized with the mean and covariance matrix computed on the whole training set by using the KL divergence. The hyperparameter is optimized with respect to the NLL on the validation set. 

\subsubsection{CIWAE}
For our CIWAE, we condition on the previous 256 days of PCA transformed factor values $\{\mathbf{\hat{F}_{t}}\}_{t=1}^{T}$ using an LSTM. In our ablation studies (\myref{Section
}{sec:ablate_factor_model}), we experiment with using the $\hat{F}_{t}$ and $\hat{F}_{E, t}$ (the EMA transformed values). We use a normal distribution for our encoder and a Normal Inverse Gaussian for our decoder; finally, we use $k=64$ with the IWAE loss. Full details on the architecture and training methodology can be found in \myref{Appendix}{sec:ciwae_arch}.

\subsection{Stock Model}\label{sec:stock_model}

Given a single stock's returns over the lookback period $\{r_{i, t}\}_{t=1}^{T}$ (analogous to GARCH) and the PCA transformed factor returns over the lookback period $\{\mathbf{\hat{F}_{t}}\}_{t=1}^{T}$, stock-level features $\mathbf{X_{i,t}}$, \textit{and} the PCA transformed factor returns over the forward-looking period $\hat{F}_{T + 1}$, we create a distribution of the stock's returns for the forward period. In essence, our stock model has to learn to forecast returns given historical data \textit{and} the next day factor values.
Mathematically, our factor model is learning:
$$ p (r_\mathbf{{i, T + 1}} | \mathbf{\hat{F}_{T + 1}},  \{r_{i, t}\}_{t=1}^{T}, \{\mathbf{\hat{F}_{t}}\}_{t=1}^{T}, \mathbf{X_{i,t}}) $$
To train our stock model, importantly, we train a single neural network across all securities and time, so the neural network learns relationships that generalize across securities and time. We ablate these feature choices in \myref{Section}{sec:ablate_stock_feat}.

\subsubsection{Conditional Flow}
Similar to the CIWAE, we use an LSTM to condition on the previous 256 days of stock returns $\{r_{i, t}\}_{t=1}^{T}$ and PCA transformed factor returns $\{\mathbf{\hat{F}_{t}}\}_{t=1}^{T}$. For the base distribution of our flow, we used a Normal Inverse Gaussian where the parameters of the distribution are outputted by a neural network; we use four conditional residual flows.
Full details on the architecture and training methodology can be found in \myref{Appendix}{sec:cflow_arch}.

\subsection{Usage}
\subsubsection{One-Day Sampling} \label{sec:one_day_sample}
Sampling from our model requires two steps: 1) sample $\mathbf{\hat{F}'_{T + 1}}$ from the factor model and then 2) sample returns for \textit{all} securities $\{r'_{i, T + 1}\}_{i=1}^{N_{T+1}}$ given $\mathbf{\hat{F}'_{T + 1}}$. Importantly, to get one $N$-dimensional sample of security returns, we use only \textit{one} sample from the factor model; this single sample induces correlation in the security returns.

\subsubsection{Multi-Day Sampling} Given a sample of factors and returns for day $T + 1$ ($\mathbf{\hat{F}'_{T + 1}}$ and $\{r'_{i, T + 1}\}_{i=1}^{N_{T+1}}$), we simply remove day $t=1$ (because we use a fixed lookback period ) and concatenate $\mathbf{\hat{F}'_{T + 1}}$ and $r'_{i, T + 1}$ to the corresponding factors time series and returns time series, respectively. We can then sample for day $T=2$ (both factor model and stock model), and repeat. Note that it is not necessary to have a fixed lookback period, but because we are using LSTMs for modeling time series, our model cannot handle lengths not seen during trainings.

\subsubsection{Statistical Quantities} Even though many important statistical quantities such as volatility are not available in closed form, because we can sample from our model, we can approximate our model's prediction of these statistical quantities via sampling. For example, for volatility, we can sample $r_{i, t}$ many times from our model and estimate the standard deviation of the samples. Similar, for correlation, we can sample $r_{i, t}$ and $r_{j, t}$ many times. 

\section{Results}\label{sec:eval}

In our experiments, we use the point-in-time constituents of the S\&P 500 as our universe, a side effect of which is that the universe can change. However, our factor analysis-inspired modeling methodology is able to handle this problem trivially.

We split our data into three pieces: the training set which contains data from the beginning of 1996 to the end of 2013, the validation set which contains data from the beginning of 2014 to the end of 2018, and the test set which contains data from the beginning of 2019 to the end of 2022. Importantly, all of our hyperparameter tuning was performed on the validation set without referring to the test set.

For evaluation, we first thoroughly ablate our different modeling decisions, specifically our choice of factors (\myref{Section}{sec:ablation_factors}), factor model architecture (\myref{Section}{sec:ablate_factor_model}), stock model features (\myref{Section}{sec:ablate_stock_feat}), and stock model architecture (\myref{Section}{sec:ablate_stock_model}).
We then compare against some classical approaches on the task of modeling the conditional joint distribution of returns (\myref{Section}{sec:baseline}) showing that our generative approach is able to outperform all of them. For ease of comparison between models, we focus on two metrics:
\begin{align} 
\text{NLL}_{\text{joint}, T}  &= \frac{1}{N_{T+1}} \log p(\{r_{i,T+1}\}_{i=1}^{N_{T+1}} | \mathcal{F}_T)
\\ &= \frac{1}{N_{T+1}} \log \int  \prod_i \left(p (r_{i, T + 1} | \mathbf{\hat{F}_{T + 1}}, \mathcal{F}_T)\right) p ( \mathbf{\hat{F}_{T + 1}} | \mathcal{F}_T) d\mathbf{\hat{F}_{T + 1}}
\end{align}
\begin{align}
\text{NLL}_{\text{ind}, T} &= \sum_{i} \frac{1}{N_{T+1}}\log p(r_{i,T} | \mathcal{F}_T)
\\ & = \sum_{i} \frac{1}{N_{T+1}} \log \int    p (r_{i, T + 1} | \mathbf{\hat{F}_{T + 1}}, \mathcal{F}_T) p ( \mathbf{\hat{F}_{T + 1}} | \mathcal{F}_T) d\mathbf{\hat{F}_{T + 1}}
\end{align}
$\text{NLL}_{\text{joint}, T}$ measures the average negative log-likelihood of the conditional joint distribution across the universe at time $T$; $\text{NLL}_{\text{ind}}$ measures the average negative log-likelihood of the conditional univariate distributions  at time $T$. $\text{NLL}_{\text{joint}}$ denotes the average $\text{NLL}_{\text{joint}, T}$ across the evaluation period, and similarly for $\text{NLL}_{\text{ind}}$.  Note that in the metrics we average the NLL for each day by the number of securities in the universe, similar to metrics used in image (bits per dim, e.g., \citet{Dinh2016NVP}) and text (perplexity, e.g., \citet{radford2019language}). For our model, we approximate these integrals using 100K samples from our factor model.

Further, we analyze qualitative behaviors of our model by analyzing commonly described stylized volatility and correlations facts in financial time series (\myref{Section}{sec:stylized}).
Finally, we evaluate the application of our model to risk analysis (\myref{Section}{sec:risk_eval}) by analyzing the calibration performance of our model on individual securities and on a simple evenly-weighted portfolio and the application of our model to portfolio optimization.

\subsection{Ablation Studies}\label{sec:ablation}

For our ablation study, we focus on the $\text{NLL}_{\text{joint}}$ on the validation set since hyperparameter tuning including model design were based on this metric.

\subsubsection{Choice of Factors}\label{sec:ablation_factors}
In \myref{Figure}{fig:pca_nll_plot}, we show the $\text{NLL}_{\text{joint}}$ given different number of factors (the lines) and different number of PCA components (the x-axis). We compare against using different sets of factors. We can see in \myref{Figure}{fig:pca_nll_plot} that the optimal model used all the factors. Further, we can see that for each factor setting, there was an optimal value of the number of PCA components. The parabolic shape that we see is to be expected as was explained in \myref{Section}{sec:factor_model}: with too few components, not enough correlation between securities is explained and with too many components, the factor distribution becomes harder to learn.

\begin{table}[!bt]
    \centering
    \small
    \begin{tabular}{lcc}
\toprule
{Model} & {Val $\text{NLL}_{\text{joint}}$}
\\
\midrule
Our Final Model  & $\mathbf{0.3932}$  \\
\midrule
\textit{Factor Model Architecture} & \\ 
VAE w/o EMA    & 0.3964   \\
EMA Gaussian  & 0.3959  \\
\midrule
\textit{Stock Model Features} & \\ 
With Industry and Financials &  0.4046  \\
No Factor Time Series    & 0.4756   \\
No Stock Time Series  & 0.5471  \\
No Time Series & 0.5481 \\
\midrule
\textit{Stock Model Architecture} & & \\
Conditional $\mathcal{NIG}$ & $0.3994$  \\
Conditional $\mathcal{N}$ & $0.7401$  \\
Linear Model & 0.6460  \\
\bottomrule
\end{tabular}
    \caption{Ablation results.}
    \label{tab:ablate}
\end{table}

\begin{figure}[!b]
    \centerline{\includegraphics[width=0.9\linewidth]{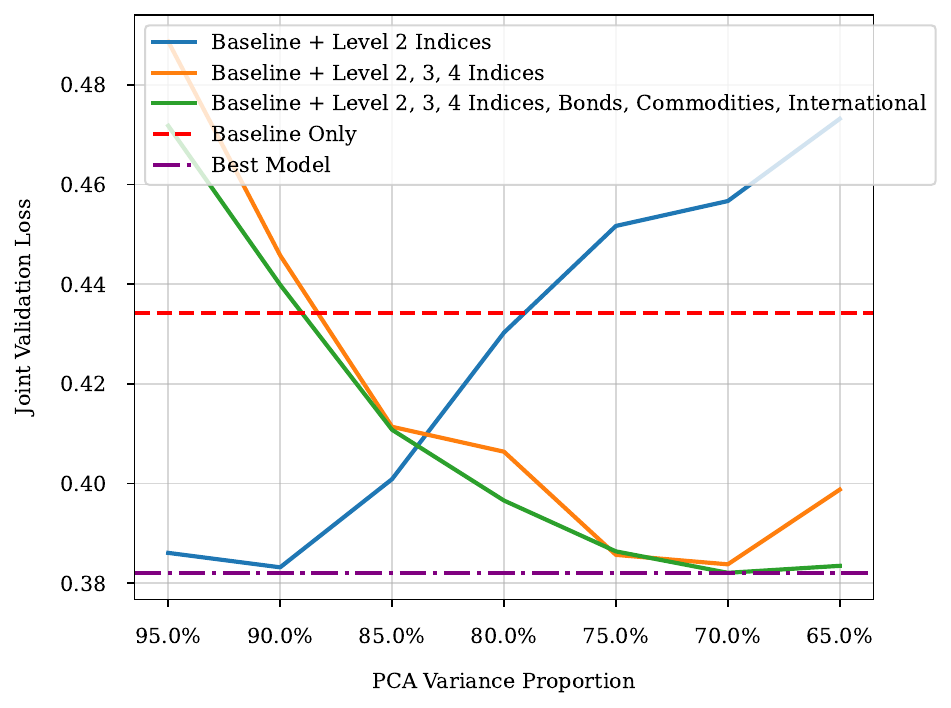}}
    \caption{Joint NLL against different number of PCA components (measured by explained variance) and different sets of factors.}
\label{fig:pca_nll_plot}
\end{figure}

\subsubsection{Factor Model Architecture} \label{sec:ablate_factor_model}
For our factor model, in \myref{Table}{tab:ablate}, we see that using a VAE outperforms using a simple multivariate Gaussian constructed using EMA; while the performance benefits do not seem much, in terms of validation likelihood for $p(\mathbf{\hat{F}_{T+ 1}} | \mathcal{F}_T)$, the validation loss improves from 0.9344 (no VAE) to 0.8794 (our final model). Further, conditioning on the EMA results in the VAE improves performance. While theoretically the VAE should be able to compute EMA features, our results suggest that directly giving the results of EMA is an effective feature engineering method. 

\subsubsection{Stock Model Features}\label{sec:ablate_stock_feat} For the feature set of our stock model, in \myref{Table}{tab:ablate}, we see that including the stock time series has the highest impact on overall performance which is to be expected because it is the stock time series that allows the model to differentiate different stocks. We see further that removing the factor time series (but still including the future factor values) has a detrimental impact. Finally, we see including industry and company financials seems to partially hurt the $\text{NLL}_{\text{joint}}$; however, we see in portfolio optimization (\myref{Section}{sec:portfolio}) that the model including industry and company financials performs the best in terms of Sharpe ratio. More work will be required to understand the source of this discrepancy.

\subsubsection{Stock Model Architecture}\label{sec:ablate_stock_model} 
For our stock model, we see in \myref{Table}{tab:ablate}
that using a normalizing flow was an important design decision, as replacing it with a simple normal or Normal Inverse Gaussian (still using a neural network but only outputting the parameters of the corresponding distribution) degrades performance. Further, using a conditional Gaussian or using only a linear model hurts performance significantly.

\subsection{Baselines}\label{sec:baseline}

For our baselines, we compared against 1) using GARCH and GJR-GARCH \citep{Glosten1993GJRGARCH} trained on each time series independently ($\text{NLL}_{\text{ind}}$) and 2) a classical factor model where we use a multivariate Gaussian on the PCA values of the factor returns using the EMA parameters from \myref{Section}{sec:factor_model} and a linear model fit per security between the factors and security returns. For the GARCH models, we set $p$, $q$ and $o$ to one. We also experimented with using a normal distribution, a generalized Gaussian (generalized error distribution, or GEM), and a skew Student's T for the noise distribution.

We see that our model is able to outperform the GARCH models and the classical factor model by {a signficiant margin}.

\begin{table}[!bt]
    \centering
    \small
    \begin{tabular}{lcccc}
\toprule
     & \multicolumn{2}{c}{Val NLL}
     & \multicolumn{2}{c}{Test NLL}
\\
 \cmidrule(lr){2-3} 
 \cmidrule(lr){4-5} 
 & {Ind} & {Joint}
 & {Ind} & {Joint}
\\
\midrule
Ours  & $\mathbf{0.728}$  & $\mathbf{0.3932}$ & $\mathbf{1.054}$  & $\mathbf{0.636}$ \\
\midrule
GARCH Normal & $0.963$  & \na & $1.299$  & \na \\
GARCH GEM & $0.860$  & \na & $1.173$  & \na \\
GARCH Skew Student & $0.856$  & \na & $1.163$  & \na \\
GJR-GARCH Skew Student & $0.858$  & \na & $1.160$  & \na \\
Classical Factor  & $0.891$  & $0.6471$ & $1.338$  & $1.004$ \\
\bottomrule
\end{tabular}
    \caption{Comparison against baseline models.}
    \label{tab:factor_model}
\end{table}

\begin{table}[!b]
    \centering
    \small
    \begin{tabular}{lcccc}
\toprule
     & \multicolumn{2}{c}{Val}
     & \multicolumn{2}{c}{Test}
\\
 \cmidrule(lr){2-3} 
 \cmidrule(lr){4-5} 
 & {ACF} & {LE}
 & {ACF} & {LE}
\\
\midrule
Ours  & $\mathbf{0.054}$  & $\mathbf{0.054}$ & $\mathbf{0.050}$  & $\mathbf{0.075}$ \\
\midrule
GARCH Skew Student & $0.060$  & $0.072$ & $0.059$  & $0.099$ \\
GJR-GARCH Skew Student & $0.062$  & $0.066$ & $0.060$  & $0.094$ \\
Classical Factor  & $0.061$  & $0.062$ & $0.062$  & $0.098$ \\
\bottomrule
\end{tabular}
    \caption{Comparison of ACF and LE against baseline models.}
    \label{tab:acf}
\end{table}

\subsection{Evaluating Stylized Volatility Facts}\label{sec:stylized}

A common use case for GARCH is as a volatility estimator. Common stylized facts that GARCH captures that makes it a strong volatility model is that it incorporates time-varying volatility and volatility clustering. 

As shown \myref{Section}{sec:baseline}, our model is able to outperform GARCH in terms of negative log-likelihood. Similar to \citet{wiese2019quantgan},  we quantitatively evaluate volatility clustering (autocorrelation in volatility, ACF) and leverage effect (LE). For ACF, per security $i$, we denote $\text{ACF}_i(d)$ as the absolute error between the correlation between $\{r_{i,t}^2\}_{t=1}^{T}$ and $\{r_{i,t - 1}^2\}_{t=1}^{T}$ and the correlation between $\{r_{i,t}^{'2}\}_{t=1}^{T}$
and $\{r_{i,t - 1}^2\}_{t=1}^{T}$ where 
$r'_{i,t}$ refers to a sampled return for security $i$ and time $t$. Similarly, for LE, we denote $\text{LE}_i(d)$ as absolute error for the correlation between $r_{i,t}^2$ and $r_{i,t - 1}$.

In \myref{Table}{tab:acf}, we use:
\begin{align*}
    \text{ACF} &= \frac{1}{3N} \sum_{i=1}^{N} \left(\text{ACF}_i(1) + \text{ACF}_i(5) + \text{ACF}_i(20)\right)
\\ \text{LE} &= \frac{1}{N} \sum_{i=1}^{N} \text{LE}_i(1)
\end{align*}
We can see that our model outperforms GARCH and the classical factor model meaning it is able to more accurately capture the autocorrelation in volatility and the leverage effect observed in historical data.

Example volatility predictions for AAPL US Equity can be seen in \myref{Appendix}{sec:volatility}.

\subsection{Risk Analysis}\label{sec:risk_eval}

\begin{table}[!b]
    \centering
    \small
    \begin{tabular}{lcccc}
\toprule
     & \multicolumn{2}{c}{Val C.E.}
     & \multicolumn{2}{c}{Test C.E.}
     \\
 \cmidrule(lr){2-3} 
 \cmidrule(lr){4-5} 
& {Uni.} &  {Port.}
& {Uni.} &  {Port.}
\\
\midrule
Ours  & $\mathbf{0.036}$  & $\mathbf{0.010}$ & $0.045$& $0.017$ \\
\midrule
GARCH Skew Student  & $0.071$ & $0.063$ & $\mathbf{0.044}$ & $\mathbf{0.006}$ \\
GJR-GARCH Skew Student  & $0.080$ & $0.069$ & $0.046$ & $0.010$ \\
Classical Factor & $0.119$ & $0.098$ & $0.143$ & $0.123$ \\
\bottomrule
\end{tabular}
    \caption{Calibration error performance. ``Uni.'' refers to taking a weighted average of calibration error per stock. ``Port.'' refers to the calibration error of an equal-weighted portfolio. }
    \label{tab:uni_ce}
\end{table}

One of the main values of having distributional estimates is in getting an estimate of the uncertainty; instead of simply getting a point estimate, often in the form of a mean, a distributional estimate allows you to understand the probability of different events, e.g., what is the probability the quantity is larger than zero. 
To evaluate the quality of our model in terms of risk, we use calibration error. 

In the context of regression, perfect calibration is defined as:
$$ \mathbb{P}(Y < \mathcal{F}_X^{-1}(p)) = p, \quad \forall p \in [0, 1] $$
where $\mathcal{F}_X$ is the predicted CDF function given $X$. Or, in words, the fraction of the data where the model CDF is less than $p$ is $p$.

\citet{kuleshov2018calibration} introduced calibration error as a metric to quantitatively measure how well the quantiles are aligned:
\begin{equation}
\begin{aligned}
 \hat{p}_j = {\abs{\{y_n  | \mathcal{F}_{x_n}(y_n) < p_j, n=1,\dots, N \}}}\ /\ {N} \\ 
 \text{cal}(y_1,\dots,y_N) = \sum_{j=1}^{M} (p_j - \hat{p}_j)^2
\end{aligned}\label{eqn:calib_err}
\end{equation}
where $\hat{p}_j$ is the fraction of the data where the model CDF is less than $p$ and $M$ is the number of quantiles that are evaluated. In this work, we set this to $100$ evenly spaced quantiles.

In \myref{Table}{tab:uni_ce}, we compute the calibration error for each stock in our validation and test set and then take the average of the calibration error across all stocks weighed by the number of days the stock is in the S\&P 500 in the corresponding period. The implication is that our model is better able to predict the quantiles of returns and better gauge risk.

In \myref{Table}{tab:uni_ce}, we compute calibration error for an equal-weighted portfolio comprised of the point-in-time stocks in the S\&P 500. We compare against  training GARCH on the historical returns of an equal-weighted portfolio. We see that our model outperforms GARCH in the validation period. Note that, while the GARCH model was trained specifically for this portfolio, our model was trained to learn the joint distribution of individual stocks; the implication is that our model was able to learn correlations effectively enough to compute accurate quantiles for portfolios. One possible implication of our model doing worse in the test period is that the correlations induced by the factors changed compared to the training and validation period or possibly a new factor was driving correlations between stocks. Additional visualization of predictions and performance for the equal-weighted portfolio can be found in \myref{Appendix}{sec:portfolio_viz}.

\subsection{Portfolio Optimization}\label{sec:portfolio}

\begin{figure}[!tb]
    \centerline{\includegraphics[width=0.9\linewidth]{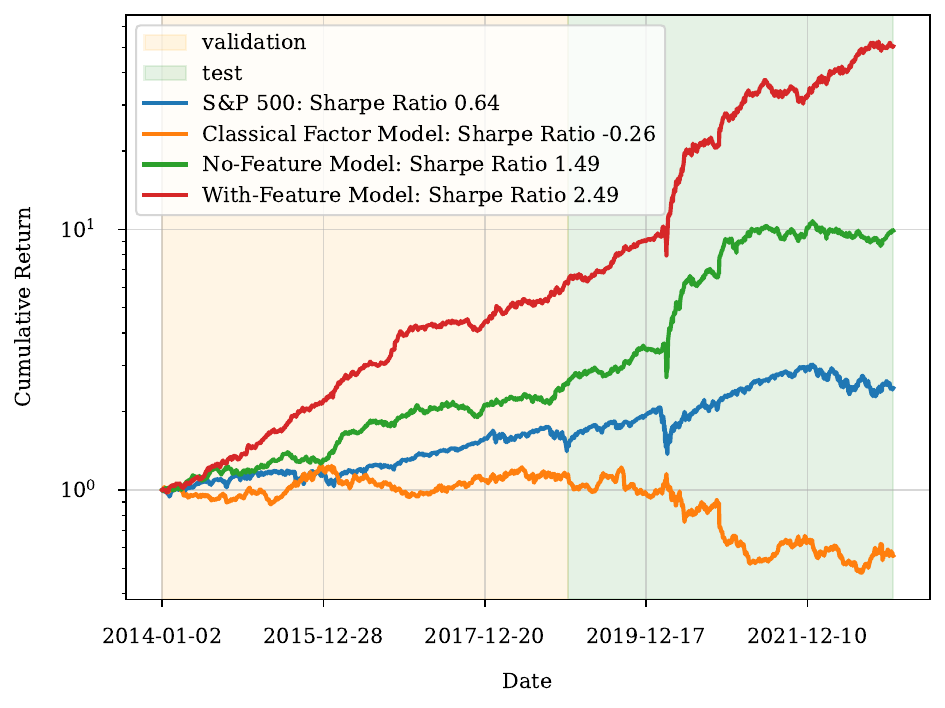}}
    \caption{Comparison of returns of long-short portfolios.}
\label{fig:port_returns}
\end{figure}

\begin{figure}[!tb]
    \centerline{\includegraphics[width=0.9\linewidth]{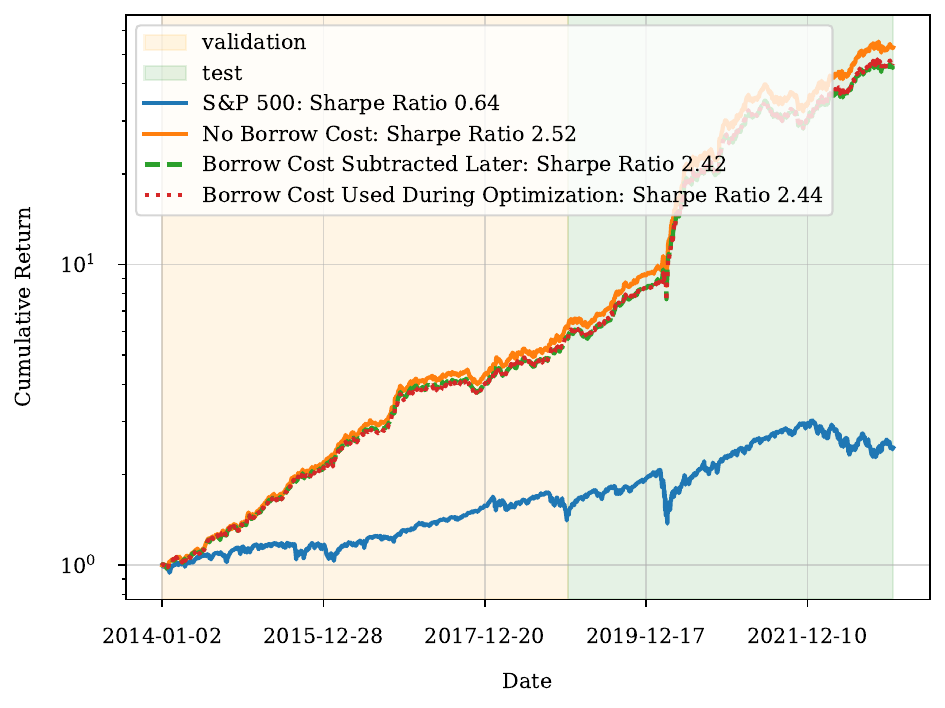}}
    \caption{Comparison of returns of long-short portfolios including borrow cost.}
\label{fig:port_returns_borrow}
\end{figure}

A common use case of volatility and correlation estimates is portfolio optimization \citep{Markowitz1952MPT}. In this section, we aim to find the long-short portfolio with the highest Sharpe ratio:
$$ \argmax_{\vec{w}_L,\vec{w}_S  \in \Delta^{N_{T+1}}} \frac{\mathbb{E}_{\vec{r}_{T+1}}\left[ \vec{w}_L^T  \vec{r}_{T+1}-\vec{w}_S^T  \vec{r}_{T+1}\right]}{\sqrt{\mathbb{V}_{\vec{r}_{T+1}}\left[ \vec{w}_L^T  \vec{r}_{T+1}-\vec{w}_S^T  \vec{r}_{T+1}\right]}} $$
where $\Delta^{N_{T+1}}$ denotes an $N_{T+1}$-dimensional simplex.

Since all the operations in the optimization task are differentiable, we used gradient descent with Adam to find the optimal long-short portfolio in the validation and test period. To make all of our final results comparable in scale to the S\&P 500, we lever up our strategy and the linear baseline to match the volatility of the S\&P 500. 

We see in \myref{Figure}{fig:port_returns} that our final model is able to outperform the classical factors model and the S\&P 500. Interestingly, however, the model including industry and company financials performs the best in terms of Sharpe ratio even though its NLL was slightly worse than without (\myref{Section}{sec:ablate_stock_feat}). {The outperformance of our model can be attributed to its minimal drawdown during Covid and quick recovery while the market was dropping.} Note, our model is trained on likelihood and is not designed to ``beat'' the market in the classical sense, but by using its well-calibrated distribution we can construct a statistically optimal portfolio. We repeat this analysis with long-only portfolios in \myref{Appendix}{sec:long_only}.

We repeat the analysis by including borrow cost in order to show a simple example where using a generative model can be used to create optimal portfolios with some additional constraints. We formulate the problem as:
$$ \argmax_{\vec{w}_L,\vec{w}_S  \in \Delta^{N_{T+1}}} \frac{\mathbb{E}_{\vec{r}_{T+1}}\left[ \vec{w}_L^T  \vec{r}_{T+1}- \vec{w}_S^T  \vec{r}_{T+1} - \vec{w}_S^T \vec{b} \right]}{\sqrt{\mathbb{V}_{\vec{r}_{T+1}}\left[ \vec{w}_L^T  \vec{r}_{T+1}-\vec{w}_S^T  \vec{r}_{T+1}\right]}} $$
where $\vec{b}$ is a vector of borrow costs for each stock. For simplicity, we set the borrow cost to $0.1$ bps, daily; we use this value since prior work (e.g., \citep{DAVOLIO2002BorrowCost}) used 17 bps per year for borrow cost which, de-annualized is approximately $0.06$ bps. We can see in \myref{Figure}{fig:port_returns_borrow} that including borrow cost for our previously optimal portfolio has a lower Sharpe ratio (2.52 to 2.42). However, when we include borrow cost into the optimization, the Sharpe ratio improves a bit.

\subsection{Analyzing the Factor Assumption}\label{sec:test_ci}
A key assumption of our approach is that all of the dependence between any pair of stocks can be explained by the ways in which both stocks depend on the factors, i.e., stock returns are conditionally independent given the stock-level features and the future factors. While our model is able to outperform other models in terms of NLL, in this section, we analyze how the correlations explained change with time.

We do so by analyzing the correlation structure in the standardized stock returns. We compute the standardized stock returns by using the mean and standard deviation as given by the stock model:
\begin{align*}
    \mu_{i,T+1} &= \mathbb{E}_{\vec{r}_{T+1} | \mathbf{F_T+1}}[\vec{r}_{i,T+1}]
\\ \sigma_{i,T+1} &= \sqrt{\mathbb{V}_{\vec{r}_{T+1}| \mathbf{F_T+1}}[\vec{r}_{i,T+1}}]
\end{align*}
We computed the correlations of standardized stock returns ($\frac{r_{i, T + 1} - \mu_{i, T+ 1}}{\sigma_{i,T+1}})$ for each calendar month in the dataset (limited to stocks that were in the index, and thus covered by the model for the entire month). We then took the root mean square value (RMS) of the upper triangle of the correlation matrix. Assuming zero correlation, \citet{Hotelling1953CorrelationCoefficient} found the exact density function of the correlation coefficient is:
$$ f(r) = \frac{(1-r^2)^{\frac{n-4}{2}}}{B\left(\frac{1}{2}, \frac{1}{2}(n-2)\right)} $$
where $B$ is the beta function and $n$ is the number of data points used to compute the correlation (i.e., in our case, one month). The RMS of the correlation matrix is equivalent to the standard deviation of the correlation coefficient:
$$ \sqrt{\frac{1}{n-1}} $$
Thus, we subtract the observed RMS by this quantity and show the performance in \myref{Figure}{fig:excess_correlation}. 

Note that when using sampled factors instead of observed factors, the standardization is performed using the covariance matrix of the resulting sampled security returns (in other words, scaled and rotated by the covariance matrix). When using observed and sampled factor values, the results are very similar, with a correlation coefficient of over 90\%.

\begin{figure}[!tb]
    \centerline{\includegraphics[width=0.9\linewidth]{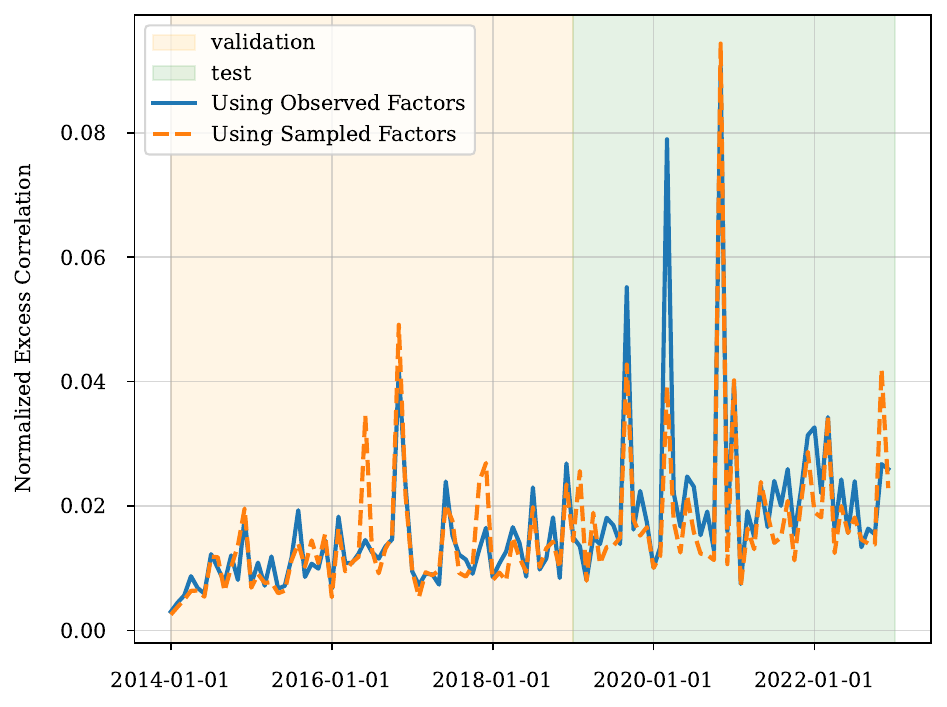}}
    \caption{Analysis of normalized excess correlation as a function of time.}
\label{fig:excess_correlation}
\end{figure}

Note that the excess correlation spikes during highly abnormal market periods like the onset of COVID lockdowns. Also note that the excess correlation rises during the testing period. This is a good indication that it is the deterioration in the joint distributional performance, specifically portfolio calibration error, that drives the worse downstream performance over this period (see \myref{Table}{tab:uni_ce}).

\subsection{The Benefit of Training on Hundreds of Stocks}\label{sec:power_of_data}

While we focus on modeling the joint distribution of hundreds of stocks in this work, a user may be interested in only a few stocks, perhaps even just one.
Nevertheless, we show in \myref{Figure}{fig:power_of_data} that training machine learning models on hundreds of stocks is beneficial even when one does not model correlations and even when one cares about the performance of the model on only one specific stock. We train a conditional flow similar to \myref{Section}{sec:stock_model} but with differing amounts of training data while always including AAPL returns in the training set. We do not include $\mathbf{\hat{F}_{T + 1}}$. For stability, we train 5 times and average the results. We can see in \myref{Figure}{fig:power_of_data} that including more stocks in the training set consistently improves the validation performance on only AAPL. Further, training on AAPL alone does not outperform GARCH. To fully realize the power of machine learning over classical financial models, it is crucial to include a large training set, even companies seemingly not in the universe of interest.

\begin{figure}[!tb]
    \centerline{\includegraphics[width=0.9\linewidth]{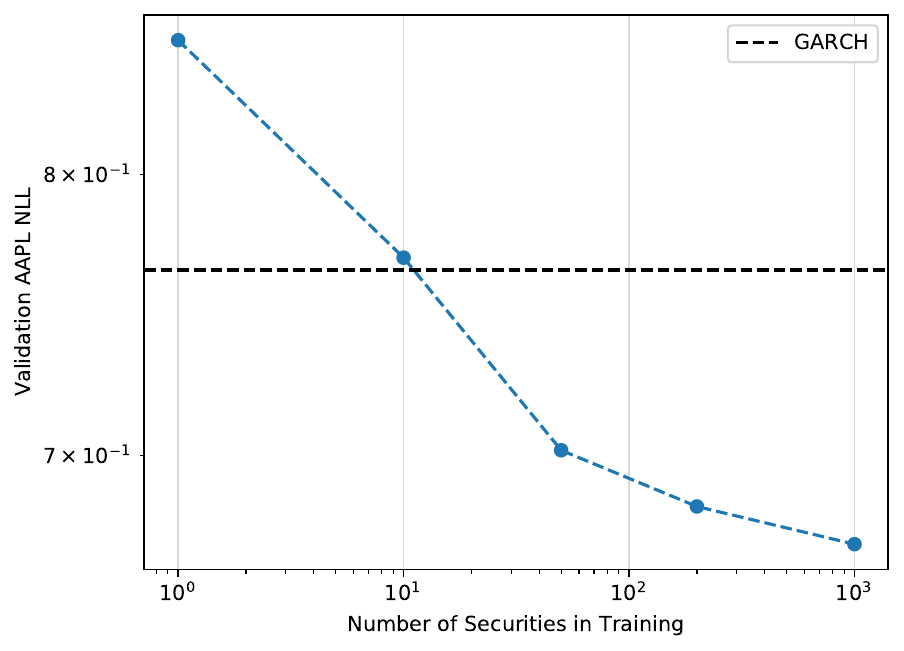}}
    \caption{NLL of Apple's returns in the validation set for differing amounts of training data, compared against GARCH.}
\label{fig:power_of_data}
\end{figure}

\section{Discussion and Conclusion}

Our framework successfully extends the concept of factor analysis into a more general approach that can be implemented via ML models. Our approach shows that ML-based distributional modeling can achieve strong results on financial tasks, and that synthetic data can be used as an input to forward-looking prediction problems. 
Notably, our model outperforms GARCH, classical factor models, and is able to create portfolios that achieve high Sharpe ratios.

Future work entails exploring additional architectures (e.g., attention \citep{vaswani2017attention}) and features (e.g., news) for the underlying ML models, using a larger universe of stocks, and doing a deeper analysis in terms of volatility and correlation estimation, as there is a deep literature for both.

With our work, we were able to provide a logical, extensible, and ML-based framework for financial problems, to demonstrate the untapped potential of ML-based approaches in finance, and to show a concrete successful 
application of synthetic data. 
While there is a large amount of work to continue doing to prove the value of generative modeling in finance, we hope the results in this paper show its potential and power.

\bibliographystyle{ACM-Reference-Format}
\bibliography{sample}

\newpage
\appendix

\section{Architecture and Training Details for CIWAE}\label{sec:ciwae_arch}

In our CIWAE architecture (\myref{Figure}{fig:vae_train_diagram}), we pass the historical EMA transformed factors (256 days) $\{\mathbf{\hat{F}_{E, t}}\}_{t=1}^{T}$ into a one-layer multi-layer perceptron (MLP) $\{\mathbf{h}_{1, t}\}_{t=1}^{T}$ and pass this set into two LSTM layers (where the first is bidirectional), giving $\mathbf{h}_{2, T}$. Intuitively, $\mathbf{h}_{2, T}$ summarizes the time-series information into a vector.

For our CIWAE, we parametrize the encoder $q(\mathbf{z} | \mathbf{\hat{F}_{E, T+1}}, \{\mathbf{\hat{F}_{E, t}}\}_{t=1}^{T})$ as:
$$ z \sim \mathcal{N}(\mu_\phi\mathbf{h_{3, T}}, \text{S}(\sigma_\phi\mathbf{h_{3, T}})) 
 \qquad \mathbf{h_{3, t}} = f_\phi(\mathbf{\hat{F}_{E, T+1}} ||\ \mathbf{h}_{2, T} )  $$
where $f_\phi$ is a neural network, $\mu_\phi, \sigma_\phi \in \RR^{\abs{h} \times \abs{z}}$, $\text{S}$ is softplus function ($\log(1 + e^x)$) to enforce positivity, and $\abs{h}$ is the hidden size. For the decoder, we use:
\begin{align*}
    \mathbf{h_{4, T}} &= g_\theta(  \mathbf{z}\ ||\ \mathbf{h}_{2, T}\ ||\ \mathbf{X}_{i, T}) \qquad \mathbf{z} \sim \mathcal{N}(0, I)
\\ \mathbf{\hat{F}_{E, T+1}} &\sim \mathcal{NIG}(
        \mu_\theta(\mathbf{h_{4, T}}),
        \text{S}(\gamma_\theta(\mathbf{h_{4, T}})),
        \beta_\theta(\mathbf{h_{4, T}}),
        \text{S}(\delta_\theta(\mathbf{h_{4, T}}))
    )
\\ \mathbf{\hat{F}_{T+1}} &=\mathbf{\Sigma_{F, T}} \mathbf{\hat{F}_{E, T+1}} + \mathbf{\mu{F, T}}
\end{align*}
where $I$ is an identity matrix, $g_\theta$ is a neural network, 
$\mu_\theta , \alpha_\theta ,\beta_\theta , \delta_\theta \in \RR^{\abs{h} \times \abs{\hat{F}}}$,  and $\mathcal{NIG}$ refers to a Normal Inverse Gaussian. Note that we use the EMA parameters to transform a Normal Inverse Gaussian and thus when computing the likelihood of $\mathbf{\hat{F}_{T+1}}$, we need to account for the change of variables induced by multiplying by $\mathbf{\Sigma_{F, T}}$. Further, while we write our decoder in terms of the sampling process, during training, we only sample from the encoder and do not need to sample from the prior or the Normal Inverse Gaussian.

All our layers used a hidden size of 128 and dropout of 0.5. We trained our factor model with an embedding size of 64 to convergence (in 50 epochs) using the Adam optimizer \citep{Adam2015} with a learning rate of 2e-4, weight decay (l2 regularization) of 2e-3, and batch size 64 with IWAE loss with $k=64$. 

For sampling (\myref{Figure}{fig:vae_gen_diagram}), we do not need to use our encoder and instead sample from the prior distribution $\mathcal{N}(0, 1)$.

\begin{figure*}
\begin{center}
\begin{tikzpicture}[
  every neuron/.style={
    circle,
    minimum size=0.3cm,
    thick
  },
  every data/.style={
    rectangle,
    minimum size=0.4cm,
    thick
  },
]

  \node [align=center,data 1/.try, minimum width=1.5cm] (input-x)  at ($ (2.5, 0) $) {$\mathbf{\hat{F}_{E, t+1}}$};

  \node [align=center,data 1/.try, minimum width=1.5cm] (input-ft)  at ($(input-x) + (-3.5,0)$) {$\mathbf{\hat{F}_{E, t}}$};
  \node [align=center,data 1/.try, minimum width=1.5cm] (input-fdots)  at ($(input-ft) + (-1.5,0)$) {$\cdots$};
  \node [align=center,data 1/.try, minimum width=1.5cm] (input-f1)  at ($(input-fdots) + (-1.5,0)$) {$\mathbf{\hat{F}_{E, 1}}$};

  \node [align=center,data 1/.try, minimum width=4.0cm, draw] (lstm)  at ($(input-fdots) + (-0.0,1.)$) {LSTM};

  \node [align=center,data 1/.try, minimum width=1.5cm] (q-mu)  at ($(input-x) + (-0.9,2.25)$) {$\mathbf{\mu_{z|\hat{F}}}$};
  \node [align=center,data 1/.try, minimum width=1.5cm] (q-sigma)  at ($(input-x) + (0.9,2.25)$) {$\mathbf{\sigma_{z|\hat{F}}}$};

  \node [align=center,data 1/.try, minimum width=1.5cm,draw] (enc-desc)  at ($(input-x) + (-.0,1.)$) {Encoder (Inference)};


  \node [align=center,data 1/.try, minimum width=1.0cm] (z)  at ($(input-x) + (0.0,3.5)$) {Sample from $\mathcal{N}(\mathbf{\mu_{z|\hat{F}}}, \mathbf{\sigma_{z|\hat{F}})}$};
  \node [align=center,data 1/.try, minimum width=1.5cm] (input-z)  at ($(enc-desc) + (0.0,3.)$) {$\mathbf{z}$};

  \node [align=center,data 1/.try, minimum width=1.5cm] (p-mu)  at ($(input-z) + (-1.5,2.25)$) {$\mathbf{\mu_{{\hat{F}}|z}}$};
  \node [align=center,data 1/.try, minimum width=1.5cm] (p-beta)  at ($(input-z) + (-0.5,2.25)$) {$\mathbf{\beta_{{\hat{F}}|z}}$};
  \node [align=center,data 1/.try, minimum width=1.5cm] (p-sigma)  at ($(input-z) + (0.5,2.25)$) {$\mathbf{\gamma_{{\hat{F}}|z}}$};
  \node [align=center,data 1/.try, minimum width=1.5cm ] (p-delta)  at ($(input-z) + (1.5,2.25)$) {$\mathbf{\delta_{{\hat{F}}|z}}$};
  \node [align=center,data 1/.try, minimum width=1.5cm, draw] (dec-desc)  at ($(input-z) + (0.0,1.)$) {Decoder  (Generator)};

  \node [align=center,data 1/.try, minimum width=1.5cm,dashed,draw] (kl)  at ($(input-z) + (4.0,0.)$) {$p(\mathbf{z}) - q(\mathbf{z} |\mathbf{\hat{F}_{E, T + 1}}, \{\mathbf{\hat{F}_{E, t}}\}_{t=1}^{T}  ) $};

\node [align=center,data 1/.try, minimum width=1.0cm, draw, dashed] (x)  at ($(input-z) + (0.0,3.5)$) {$p(\mathbf{\hat{F}_{E, t+1}} |  \mathcal{NIG}(
        \mu_{{\hat{F}}|z},
        \gamma_{{\hat{F}}|z},
        \beta_{{\hat{F}}|z},
        \delta_{{\hat{F}}|z})$};

    \draw [black,solid,->] ($(input-f1.north)$) -- ($(input-f1.north) + (0,0.5)$);
    \draw [black,solid,->] ($(input-ft.north)$) -- ($(input-ft.north) + (0,0.5)$);
  \draw [black,solid,->] ($(lstm.east)$) -- ($(enc-desc.west)$);
  \draw [black,solid,->] ($(lstm.east)$) -- ($(dec-desc.west)$);
  \draw [black,solid,->] ($(input-x.north)$) -- ($(enc-desc.south)$);
  \draw [black,solid,->] ($(enc-desc.north)$) -- ($(q-mu.south)$);
  \draw [black,solid,->] ($(enc-desc.north)$) -- ($(q-sigma.south)$);

  \draw [black,solid,->] ($(q-mu.north)$) -- ($(z.south)$);
  \draw [black,solid,->] ($(q-sigma.north)$) -- ($(z.south)$);

  \draw [black,solid,->] ($(input-z.north)$) -- ($(dec-desc.south)$);
  
  \draw [black,solid,->] ($(dec-desc.north)$) -- ($(p-beta.south)$);
  \draw [black,solid,->] ($(dec-desc.north)$) -- ($(p-delta.south)$);
  \draw [black,solid,->] ($(dec-desc.north)$) -- ($(p-mu.south)$);
  \draw [black,solid,->] ($(dec-desc.north)$) -- ($(p-sigma.south)$);

  \draw [black,solid,->] ($(p-mu.north)$) -- ($(x.south)$);
  \draw [black,solid,->] ($(p-sigma.north)$) -- ($(x.south)$);
  \draw [black,solid,->] ($(p-beta.north)$) -- ($(x.south)$);
  \draw [black,solid,->] ($(p-delta.north)$) -- ($(x.south)$);

\end{tikzpicture}
\end{center}
\caption{A diagrammatic representation of training our CIWAE. The loss terms are in dashed boxes. }\label{fig:vae_train_diagram}
\end{figure*}
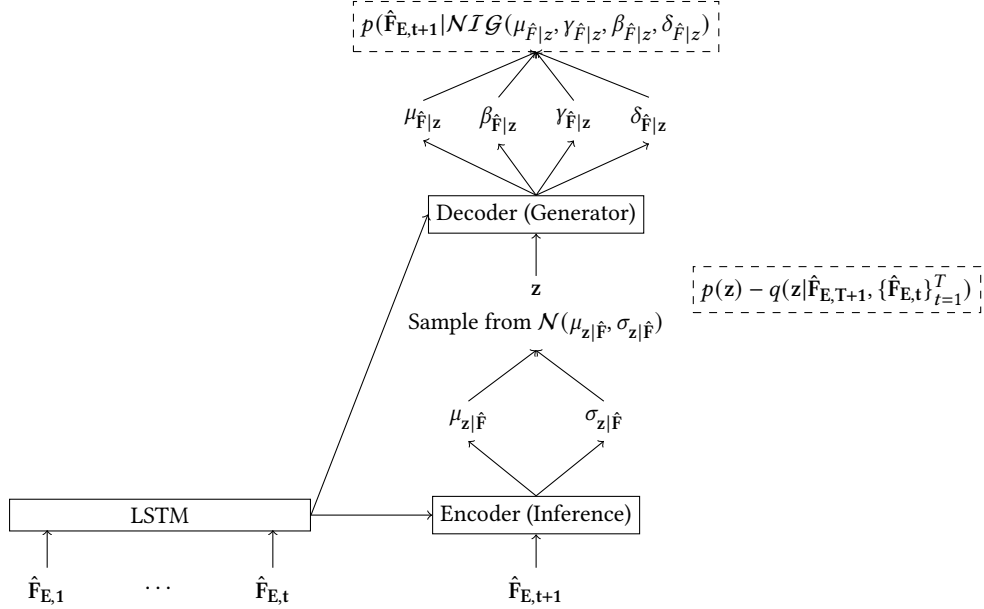

\begin{figure*}
\begin{center}
\begin{tikzpicture}[
  every neuron/.style={
    circle,
    minimum size=0.3cm,
    thick
  },
  every data/.style={
    rectangle,
    minimum size=0.4cm,
    thick
  },
]

  \node [align=center,data 1/.try, minimum width=1.5cm] (input-z)  at ($ (2.5, 0) $) {$\mathbf{z}$};
  \node [align=center,data 1/.try, minimum width=1.0cm] (z)  at ($(input-z) - (0.0,0.25)$) {Sample from $\mathcal{N}(0,1)$};

  \node [align=center,data 1/.try, minimum width=1.5cm] (input-ft)  at ($(input-x) + (-3.5,0)$) {$\mathbf{\hat{F}_{E, t}}$};
  \node [align=center,data 1/.try, minimum width=1.5cm] (input-fdots)  at ($(input-ft) + (-1.5,0)$) {$\cdots$};
  \node [align=center,data 1/.try, minimum width=1.5cm] (input-f1)  at ($(input-fdots) + (-1.5,0)$) {$\mathbf{\hat{F}_{E, 1}}$};

  \node [align=center,data 1/.try, minimum width=4.0cm, draw] (lstm)  at ($(input-fdots) + (-0.0,1.)$) {LSTM};


  \node [align=center,data 1/.try, minimum width=1.5cm] (p-mu)  at ($(input-z) + (-1.5,2.25)$) {$\mathbf{\mu_{{\hat{F}}|z}}$};
  \node [align=center,data 1/.try, minimum width=1.5cm] (p-beta)  at ($(input-z) + (-0.5,2.25)$) {$\mathbf{\beta_{{\hat{F}}|z}}$};
  \node [align=center,data 1/.try, minimum width=1.5cm] (p-sigma)  at ($(input-z) + (0.5,2.25)$) {$\mathbf{\gamma_{{\hat{F}}|z}}$};
  \node [align=center,data 1/.try, minimum width=1.5cm ] (p-delta)  at ($(input-z) + (1.5,2.25)$) {$\mathbf{\delta_{{\hat{F}}|z}}$};
  \node [align=center,data 1/.try, minimum width=1.5cm,draw] (dec-desc)  at ($(input-z) + (0.0,1.)$) {Decoder  (Generator)};

\node [align=center,data 1/.try, minimum width=1.0cm] (x)  at ($(input-z) + (0.0,3.5)$) {Sample from $\mathcal{NIG}(
        \mu_{{\hat{F}}|z},
        \gamma_{{\hat{F}}|z},
        \beta_{{\hat{F}}|z},
        \delta_{{\hat{F}}|z})$};
  \node [align=center,data 1/.try, minimum width=1.5cm] (output-x)  at ($(dec-desc) + (0.0,3.2)$) {$\mathbf{\hat{F}_{E, t+1}}$};

    \draw [black,solid,->] ($(input-f1.north)$) -- ($(input-f1.north) + (0,0.5)$);
    \draw [black,solid,->] ($(input-ft.north)$) -- ($(input-ft.north) + (0,0.5)$);
  \draw [black,solid,->] ($(lstm.east)$) -- ($(dec-desc.west)$);


  \draw [black,solid,->] ($(input-z.north)$) -- ($(dec-desc.south)$);
  
  \draw [black,solid,->] ($(dec-desc.north)$) -- ($(p-beta.south)$);
  \draw [black,solid,->] ($(dec-desc.north)$) -- ($(p-delta.south)$);
  \draw [black,solid,->] ($(dec-desc.north)$) -- ($(p-mu.south)$);
  \draw [black,solid,->] ($(dec-desc.north)$) -- ($(p-sigma.south)$);

  \draw [black,solid,->] ($(p-mu.north)$) -- ($(x.south)$);
  \draw [black,solid,->] ($(p-sigma.north)$) -- ($(x.south)$);
  \draw [black,solid,->] ($(p-beta.north)$) -- ($(x.south)$);
  \draw [black,solid,->] ($(p-delta.north)$) -- ($(x.south)$);

\end{tikzpicture}
\end{center}
\caption{A diagrammatic representation of sampling from our CIWAE.  }\label{fig:vae_gen_diagram}
\end{figure*}
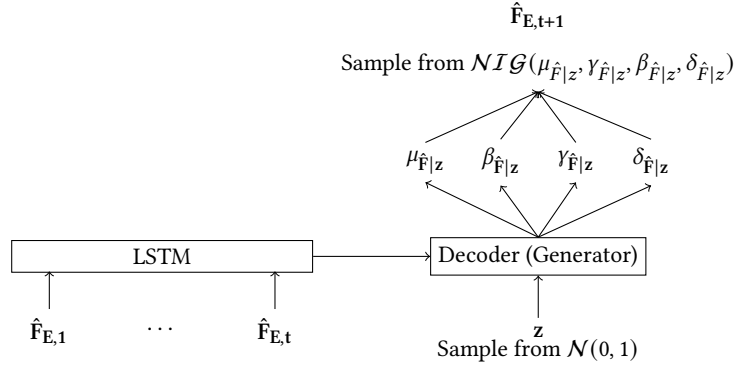

\section{Architecture and Training Details of Flow}\label{sec:cflow_arch}

For our conditional flow architecture (\myref{Figure}{fig:cflow_train_diagram}),  we concatenate the stock return $\{r_{i, t}\}_{t=1}^{T}$ and PCA transformed factor returns $\{\mathbf{\hat{F}_{t}}\}_{t=1}^{T}$ (256 days) and, similar to the factor model, pass this into a one-layer multi-layer perceptron (MLP) $\{\mathbf{h}_{1, i, t}\}_{t=1}^{T}$  and then into two LSTM layers (where the first is bidirectional), giving $\mathbf{h}_{2, i,T}$.

$\mathbf{h}_{2, i,T}$ is concatenated with $\mathbf{\hat{F}_{T + 1}}$ and is  passed through another two-layer MLP $\mathbf{h}_{3, i,T}$. The normalizing flow is then constructed:
\begin{align}
    z &\sim \mathcal{NIG}(
        \mathbf{\mu_\theta}^T\mathbf{h_{3,i, T}},
        \text{S}(\mathbf{\gamma_\theta}^T\mathbf{h_{3, i, T}}),
        \mathbf{\beta_\theta}^T\mathbf{h_{3, i, T}},
        \text{S}(\mathbf{\delta_\theta}^T\mathbf{h_{3, i, T}})
    )
\\ r_{i, t} &= g_\theta(z, \mathbf{h_{3, T}})
\end{align}
where $g_\theta$ is the composition of four conditional residual flows.
Note, because $r_{i, T+1}$ is a scalar, the issue of computing the log-determinant for the loss is moot; we can compute $\abs{\frac{\partial f(x)}{\partial x}}$ using automatic differentiation.
Further, in the one-dimensional case, we can trivially compute the CDF function: $\phi_z(f(x))$ where $\phi_z$ is the CDF of the base distribution. This was used during calibration evaluation (\myref{Section}{sec:risk_eval}).

All our layers used a hidden size of 256 and dropout of 0.5. We trained our stock model for 200K gradient steps using the Adam optimizer \citep{Adam2015} with a learning rate of 1e-3, weight decay (l2 regularization) of 2e-2, and batch size 128. Note, that by design of the architecture, during sampling (\myref{Figure}{fig:cflow_gen_diagram}), we can precompute $\mathbf{h}_{2, i}$ as it is independent of future factor values.

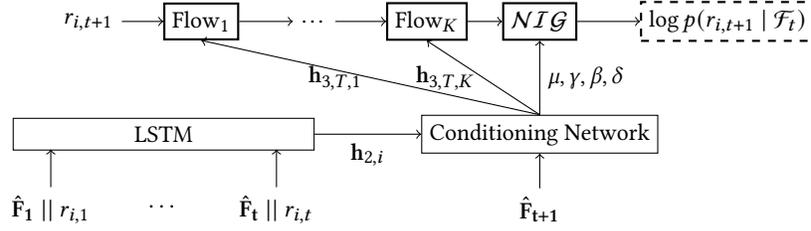
\begin{figure*}
\begin{center}
\begin{tikzpicture}[
  every neuron/.style={
    circle,
    minimum size=0.3cm,
    thick
  },
  every data/.style={
    rectangle,
    minimum size=0.4cm,
    thick
  },
]

  \node [align=center,data 1/.try, minimum width=1.5cm] (input-x)  at ($ (2.5, 0) $) {$\mathbf{\hat{F}_{t+1}}$};

  \node [align=center,data 1/.try, minimum width=1.5cm] (input-ft)  at ($(input-x) + (-3.5,0)$) {$\mathbf{\hat{F}_{t}}\ ||\ r_{i,t}$};
  \node [align=center,data 1/.try, minimum width=1.5cm] (input-fdots)  at ($(input-ft) + (-1.5,0)$) {$\cdots$};
  \node [align=center,data 1/.try, minimum width=1.5cm] (input-f1)  at ($(input-fdots) + (-1.5,0)$) {$\mathbf{\hat{F}_{1}}\ ||\  r_{i,1}$};

  \node [align=center,data 1/.try, minimum width=4.0cm, draw] (lstm)  at ($(input-fdots) + (-0.0,1.)$) {LSTM};

  \node [align=center,data 1/.try, minimum width=1.5cm,draw] (mlp-4)  at ($(input-x) + (-.0,1.)$) {Conditioning Network};

    \draw [black,solid,->] ($(input-f1.north)$) -- ($(input-f1.north) + (0,0.5)$);
    \draw [black,solid,->] ($(input-ft.north)$) -- ($(input-ft.north) + (0,0.5)$);
  \draw [black,solid,->] ($(lstm.east)$) -- ($(mlp-4.west)$) node [midway,below] {$\mathbf{h}_{2, i}$};
  \draw [black,solid,->] ($(input-x.north)$) -- ($(mlp-4.south)$);

  \node [align=center,every data/.try, data 1/.try, minimum height=0.4cm,draw] (flow-1)  at ($ (mlp-4) + (-4.5,1.5) $) {$\text{Flow}_1$};

  \node [align=center,every data/.try, data 1/.try, minimum height=0.4cm] (target)  at ($ (flow-1) - (1.5,0) $) {$r_{i, t+1}$};

  \node [align=center,every data/.try, data 1/.try, minimum height=0.4cm] (flow-3)  at ($ (flow-1) + (1.5,0) $) {\dots};
  \node [align=center,every data/.try, data 1/.try, minimum height=0.4cm,draw] (flow-4)  at ($ (flow-3) + (1.5,0) $) {$\text{Flow}_K$};
  \node [align=center,every data/.try, data 1/.try, minimum height=0.4cm,draw] (base)  at ($ (flow-4) + (1.5,0) $) {$\mathcal{NIG}$};

  \node [align=center,every data/.try, data 1/.try, minimum height=0.4cm, draw,dashed] (loss)  at ($ (base) + (2.5,0) $) {$\log p(r_{i, t+1}\ |\ \mathcal{F}_t)$};

  \draw [black,solid,->] ($(target.east)$) -- ($(flow-1.west)$);
  \draw [black,solid,->] ($(flow-1.east)$) -- ($(flow-3.west)$);
  \draw [black,solid,->] ($(flow-3.east)$) -- ($(flow-4.west)$);
  \draw [black,solid,->] ($(flow-4.east)$) -- ($(base.west)$);
  \draw [black,solid,->] ($(base.east)$) -- ($(loss.west)$);

    \draw [black,solid,->] ($(mlp-4.north)$) -- ($(flow-1.south)$) node[midway,left]  {$\mathbf{h}_{3,T,1}$};
  \draw [black,solid,->] ($(mlp-4.north)$) -- ($(flow-4.south)$) node[midway,left]  {$\mathbf{h}_{3,T,K}$};
  \draw [black,solid,->] ($(mlp-4.north)$) -- ($(base.south)$) node[midway,right]  {$\mathbf{\mu},\mathbf{\gamma},\mathbf{\beta},\mathbf{\delta}$};

\end{tikzpicture}
\end{center}
\caption{A diagrammatic representation of training  our conditional flow.}\label{fig:cflow_train_diagram}
\end{figure*}

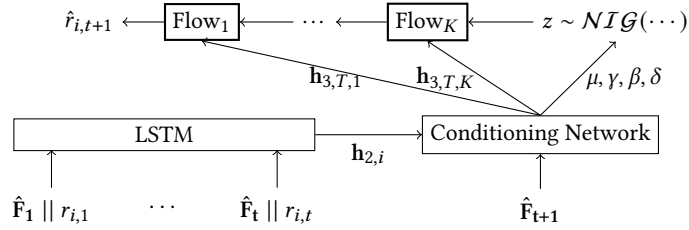
\begin{figure*}
\begin{center}
\begin{tikzpicture}[
  every neuron/.style={
    circle,
    minimum size=0.3cm,
    thick
  },
  every data/.style={
    rectangle,
    minimum size=0.4cm,
    thick
  },
]

  \node [align=center,data 1/.try, minimum width=1.5cm] (input-x)  at ($ (2.5, 0) $) {$\mathbf{\hat{F}_{t+1}}$};

  \node [align=center,data 1/.try, minimum width=1.5cm] (input-ft)  at ($(input-x) + (-3.5,0)$) {$\mathbf{\hat{F}_{t}}\ ||\ r_{i,t}$};
  \node [align=center,data 1/.try, minimum width=1.5cm] (input-fdots)  at ($(input-ft) + (-1.5,0)$) {$\cdots$};
  \node [align=center,data 1/.try, minimum width=1.5cm] (input-f1)  at ($(input-fdots) + (-1.5,0)$) {$\mathbf{\hat{F}_{1}}\ ||\  r_{i,1}$};

  \node [align=center,data 1/.try, minimum width=4.0cm, draw] (lstm)  at ($(input-fdots) + (-0.0,1.)$) {LSTM};

  \node [align=center,data 1/.try, minimum width=1.5cm,draw] (mlp-4)  at ($(input-x) + (-.0,1.)$) {Conditioning Network};

    \draw [black,solid,->] ($(input-f1.north)$) -- ($(input-f1.north) + (0,0.5)$);
    \draw [black,solid,->] ($(input-ft.north)$) -- ($(input-ft.north) + (0,0.5)$);
  \draw [black,solid,->] ($(lstm.east)$) -- ($(mlp-4.west)$) node [midway,below] {$\mathbf{h}_{2, i}$};
  \draw [black,solid,->] ($(input-x.north)$) -- ($(mlp-4.south)$);

  \node [align=center,every data/.try, data 1/.try, minimum height=0.4cm,draw] (flow-1)  at ($ (mlp-4) + (-4.5,1.5) $) {$\text{Flow}_1$};

  \node [align=center,every data/.try, data 1/.try, minimum height=0.4cm] (target)  at ($ (flow-1) - (1.5,0) $) {$\hat{r}_{i, t+1}$};

  \node [align=center,every data/.try, data 1/.try, minimum height=0.4cm] (flow-3)  at ($ (flow-1) + (1.5,0) $) {\dots};
  \node [align=center,every data/.try, data 1/.try, minimum height=0.4cm,draw] (flow-4)  at ($ (flow-3) + (1.5,0) $) {$\text{Flow}_K$};
  \node [align=center,every data/.try, data 1/.try, minimum height=0.4cm] (base)  at ($ (flow-4) + (2.5,0) $) {$z \sim \mathcal{NIG}(\cdots)$};

  \draw [black,solid,<-] ($(target.east)$) -- ($(flow-1.west)$);
  \draw [black,solid,<-] ($(flow-1.east)$) -- ($(flow-3.west)$);
  \draw [black,solid,<-] ($(flow-3.east)$) -- ($(flow-4.west)$);
  \draw [black,solid,<-] ($(flow-4.east)$) -- ($(base.west)$);

    \draw [black,solid,->] ($(mlp-4.north)$) -- ($(flow-1.south)$) node[midway,left]  {$\mathbf{h}_{3,T,1}$};
  \draw [black,solid,->] ($(mlp-4.north)$) -- ($(flow-4.south)$) node[midway,left]  {$\mathbf{h}_{3,T,K}$};
  \draw [black,solid,->] ($(mlp-4.north)$) -- ($(base.south)$) node[midway,right]  {$\mathbf{\mu},\mathbf{\gamma},\mathbf{\beta},\mathbf{\delta}$};

\end{tikzpicture}
\end{center}
\caption{A diagrammatic representation of sampling from our conditional flow.}\label{fig:cflow_gen_diagram}
\end{figure*}

\section{Long-Only Portfolio Optimization}\label{sec:long_only}

For long-only portfolios with the highest Sharpe ratio, we optimize:
$$ \argmax_{\vec{w}_L \in \Delta^{N_{T+1}}} \frac{\mathbb{E}_{\vec{r}_{T+1}}\left[ \vec{w}_L^T  \vec{r}_{T+1}\right]}{\sqrt{\mathbb{V}_{\vec{r}_{T+1}}\left[ \vec{w}_L^T  \vec{r}_{T+1}\right]}} $$
where $\Delta^{N_{T+1}}$ denotes an $N_{T+1}$-dimensional simplex. In \myref{Figure}{fig:port_returns_longonly}, our model outperforms the market and the classical factor model; whereas in the long-short portfolio, we found our model with industry and company financials had the best Sharpe ratio, for long-only portfolios, we find that our model with the best $\text{NLL}_{\text{joint}}$ has the best Sharpe ratio.

\begin{figure}[!b]
    \centerline{\includegraphics[width=0.9\linewidth]{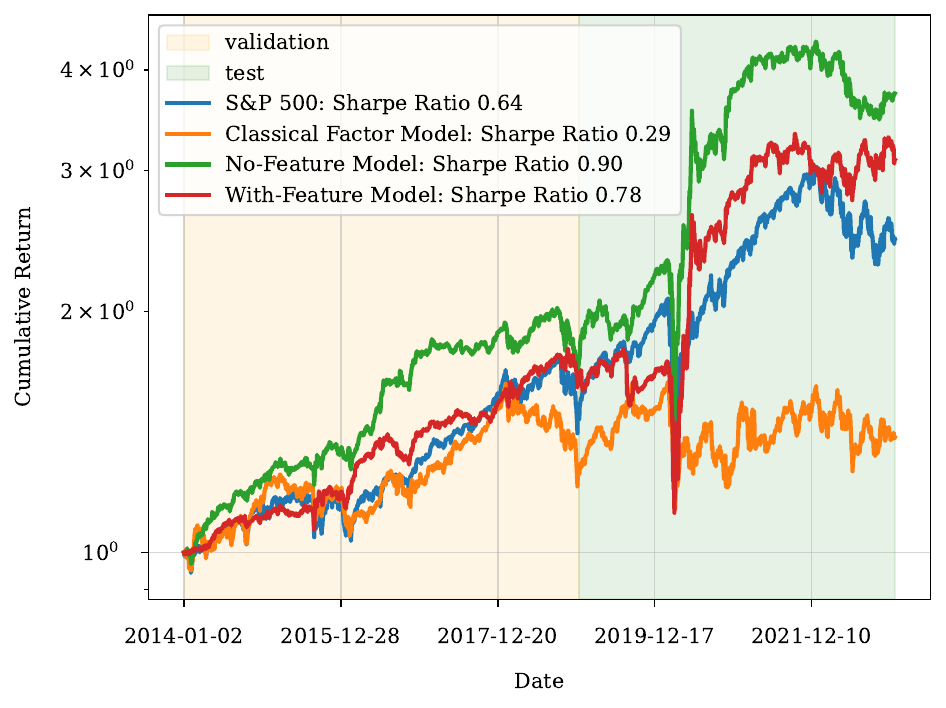}}
    \caption{Comparison of returns of long-only portfolios.}
\label{fig:port_returns_longonly}
\end{figure}

\section{Volatility Example}\label{sec:volatility}

In \myref{Figure}{fig:vol_aapl}, we show an example of volatilites predicted by our model compared to GARCH. We see that the RMSE between the observed squared returns and our model is less than the RMSE against GARCH.
\begin{figure}[!b]
    \centerline{\includegraphics[width=0.9\linewidth]{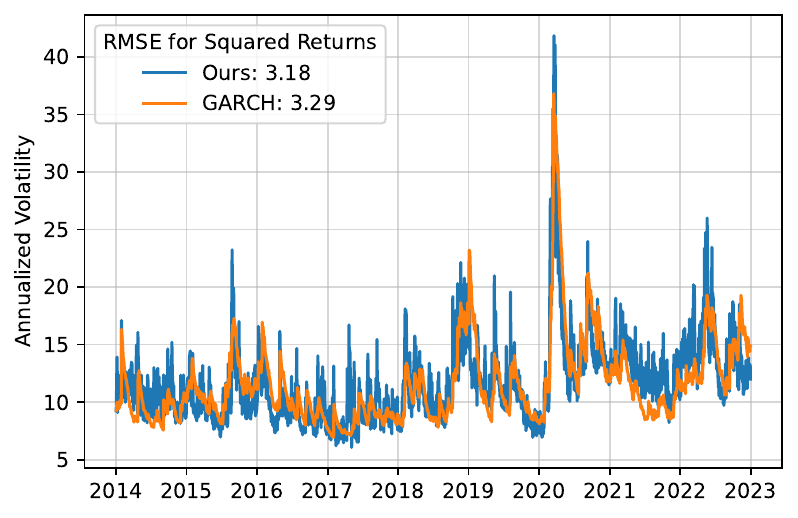}}
    \caption{Comparison of volatility predictions by our model and GARCH.}
\label{fig:vol_aapl}
\end{figure}

\section{Portfolio Returns Visualization}\label{sec:portfolio_viz}

\begin{figure}[!bh]
    \centerline{\includegraphics[width=0.9\linewidth]{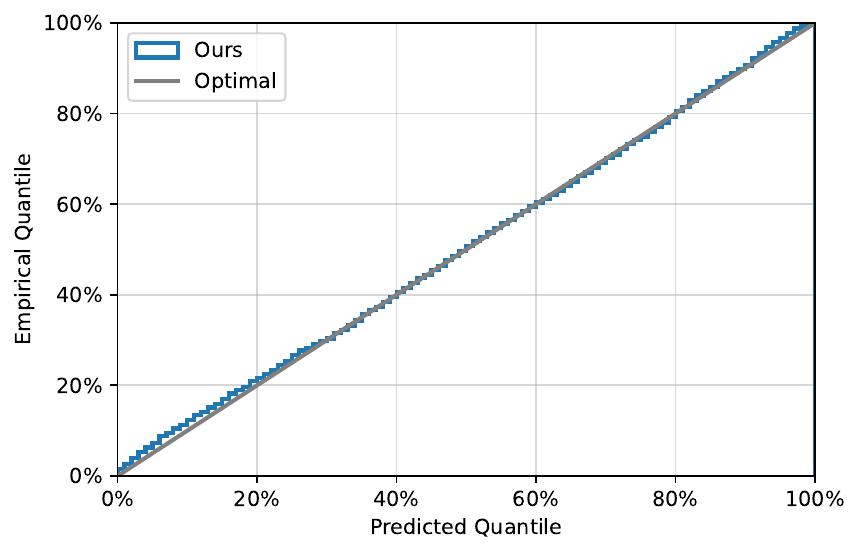}}
    \caption{Q-Q plot comparing quantiles predicted by our model compared to the empirical quantiles for an equal-weighted portfolio comprised of the point-in-time stocks in the S\&P 500 in validation and test period.}
\label{fig:portfolio_qq_plot}
\end{figure}

\begin{figure}[!bh]
    \centerline{\includegraphics[width=0.9\linewidth]{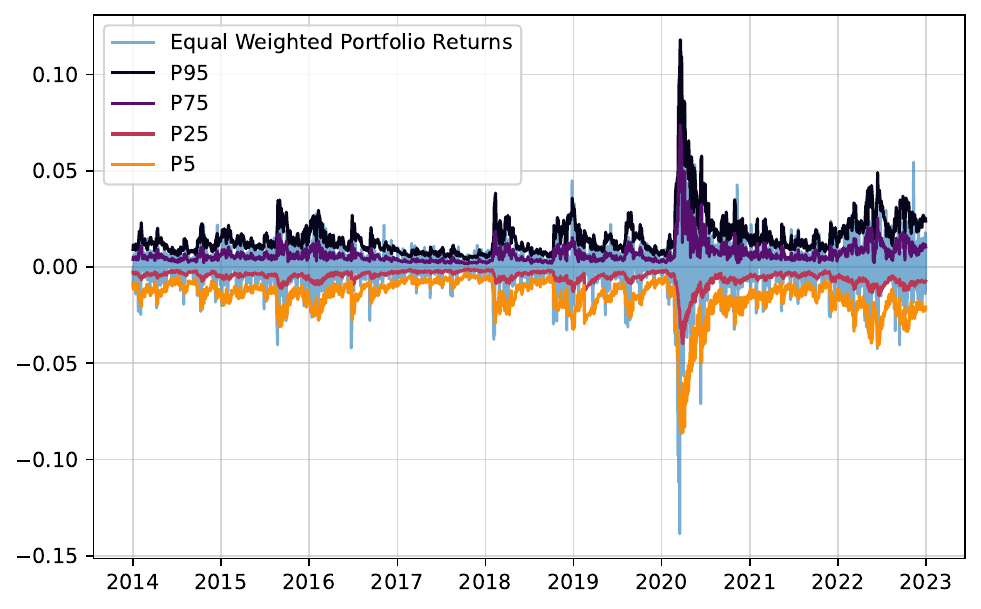}}
    \caption{Plot shows the returns of an equal-weighted portfolio comprised of the point-in-time stocks and the predicted 5th, 25th, 75th, and 95th quantile.}
\label{fig:portfolio_quantile_pred}
\end{figure}

\end{document}